\documentstyle[preprint,aps,psfig]{revtex}
\def\TL{\hfil$\displaystyle{##}$}
\def\TR{$\displaystyle{{}##}$\hfil}
\def\TC{\hfil$\displaystyle{##}$\hfil}
\def\TT{\hbox{##}}
\def\seqalign#1#2{\vcenter{\openup1\jot
  \halign{\strut #1\cr #2 \cr}}}

\def\comment#1{}
\def\fixit#1{}

\def\tf#1#2{{\textstyle{#1 \over #2}}}

\def\mop#1{\mathop{\rm #1}\nolimits}

\def\tr{\mop{tr}}

\def\overleftrightarrow#1{\vbox{\ialign{##\crcr
     $\leftrightarrow$\crcr\noalign{\kern-0pt\nointerlineskip}
     $\hfil\displaystyle{#1}\hfil$\crcr}}}

\def\lsim{\mathrel{\mathstrut\smash{\ooalign{\raise2.5pt\hbox{$<$}\cr\lower2.5pt\hbox{$\sim$}}}}}
\def\gsim{\mathrel{\mathstrut\smash{\ooalign{\raise2.5pt\hbox{$>$}\cr\lower2.5pt\hbox{$\sim$}}}}}

\def\sqr#1#2{{\vcenter{\vbox{\hrule height.#2pt
         \hbox{\vrule width.#2pt height#1pt \kern#1pt
            \vrule width.#2pt}
         \hrule height.#2pt}}}}

\def\href#1#2{#2}  

\def\lbldef#1#2{\expandafter\gdef\csname #1\endcsname {#2}}
\def\eqn#1#2{\lbldef{#1}{(\ref{#1})}%
\begin{equation} #2 \label{#1} \end{equation}}
\def\eqalign#1{\vcenter{\openup1\jot
    \halign{\strut\span\TL & \span\TR\cr #1 \cr
   }}}
\def\eno#1{(\ref{#1})}

\def\Li{\mop{Li}}
\begin{document}
\baselineskip=15.6pt
\renewcommand{\baselinestretch}{1}  
\pagestyle{plain}
\setcounter{page}{1}
\renewcommand{\thefootnote}{\fnsymbol{footnote}}

\preprint{\vbox{\baselineskip=12pt
\rightline{HUTP-{\sl A}086}
\vskip0.2truecm
\rightline{UPR-826-T}
\vskip0.2truecm
\rightline{hep-th/9903132}}}

\title{Thermodynamic Stability  and Phases of General Spinning Branes}
\author{Mirjam 
Cveti\v{c}$^{(1)}$,  and Steven S.~Gubser$^{(2)}$} 
\address{
(1)  David  Rittenhouse  Laboratories, 
University  of  Pennsylvania,
Philadelphia,  PA  19104 
\\
(2) Lyman Laboratory of Physics, Harvard University,
Cambridge, MA 02138}

\maketitle
\begin{abstract}
 We determine the thermodynamic stability conditions for near-extreme
rotating D3, M5, and M2-branes with multiple angular momenta.  Critical
exponents near the boundary of stability are discussed and compared with a
naive field theory model.  From a partially numerical computation we
conclude that outside the boundary of stability, the angular momentum
density tends to become spatially inhomogeneous.

Periodic Euclidean spinning brane solutions have been studied as models of
QCD.  We explain how supersymmetry is restored in the world-volume field
theory in a limit where spin becomes large compared to total energy.  We
discuss the hierarchy of energy scales that develops as this limit is
approached.

\end{abstract}
\renewcommand{\baselinestretch}{0.6}  
\newpage
\section{Introduction}

Mathematically, thermodynamic stability is equated with the
subadditivity of the entropy function.  More precisely, suppose the
entropy $S$ is known in the microcanonical ensemble as a function of
the other extensive thermodynamic variables (like energy and charge),
which we will collectively denote $x_i$.  Subadditivity of the
function $S(\{x_i\})$ is the condition
  \eqn{SubAdd}{
   S(\{\lambda x_i + (1-\lambda) \tilde{x}_i\}) \geq
    \lambda S(\{x_i\}) + (1-\lambda) S(\{\tilde{x}_i\}) \ .
  }
 If the inequality went the other way, then the system could gain
entropy by dividing into two parts, one with a fraction $\lambda$ of
the energy, charge, etc., and the other with a fraction $1-\lambda$.
When the system satisfies \SubAdd, no such process is allowed by the
Second Law of Thermodynamics, and it is in this sense that the system
is thermodynamically stable.

For black holes, perhaps a better way to think of \SubAdd\ (though a more
heuristic one) is as follows: \SubAdd\ is satisfied for all $x_i$ and
$\tilde{x}_i$ in a sufficiently small neighborhood of a given point in
phase space if and only if it is possible for a black hole at that point to
exist in thermal equilibrium with a heat bath much larger than itself.  A
simple example would be the Reissner-Nordstr\"om black hole in
four-dimensional Einstein-Maxwell theory with no electrically charged
particles.  Near extremality, the specific heat is positive, but far from
extremality the specific heat is negative.  If one starts with the black
hole in a state with positive specific heat and a temperature close to the
temperature of some thermal gas of gravitons and photons, then the black
hole will equilibrate with the thermal gas.  On the other hand, if the
black hole was in a state of negative specific heat, then even if its
temperature is arbitrarily close to that of the thermal gas, the black hole
will either emit faster than it absorbs or absorb faster than it emits, and
correspondingly get smaller and hotter or bigger and colder.  There is no
bound on the runaway phenomenon except for the black hole to make the
transition to positive specific heat or else to absorb most of the matter
in the universe (assuming the universe is finite).  That is what one means
by instability of the initial configuration.\footnote{In this discussion we
have assumed that the Jean's instability of the thermal gas is
insignificant---that is, it enters into the problem with a much longer
time-scale than the time scale of equilibration between the black hole and
the thermal gas.}

For spinning D- or M-branes, the heat bath line of reasoning is
questionable because it is not clear whether one can construct an
appropriate heat bath.  One could perform the following gedanken
experiment.  Start with an array of nearly identical parallel spinning
branes distributed through space, and ask whether, as they trade
Hawking radiation among themselves, there is a process of
equilibration or disequilibration among them.  This is the standard
trick of constructing an ensemble out of replicas of a given system.
The branes should equilibrate precisely when \SubAdd\ is satisfied.
However, there is a difficulty: one has to estimate whether the time
scale of equilibration or disequilibration is faster or slower than
the time scale on which the gravitational forces between the replicas
appreciably distort the array.  Near the phase transition of
\cite{gspin}, one expects the equilibration time scale to diverge, so
the gedanken experiment will become impractical (even by the standards
of gedanken experiments) close enough to the transition.

As a first step we can think of a single cluster of near-extreme
coincident branes in a decoupling limit of the theory, where Hawking
radiation to the asymptotically flat region of spacetime is very slow.
Once we have analyzed the criterion \SubAdd\ and established the
existence of unstable solutions (even in this near-extreme decoupling
limit), we can ask what form the instability takes.  Because the
branes have spatial extent, perhaps the most natural assumption is
that the solutions which spin too fast are unstable against
distributing their angular momentum unevenly across their
worldvolume, somewhat analogous to the Gregory-Laflamme transition
\cite{gl}.  We will have more to say on this possibility in
section~\ref{PhaseMix}.

The work of \cite{gspin} illustrates that the thermodynamics of the
D3-brane world-volume theory, as read off from the Bekenstein-Hawking
formula from the supergravity geometry, is stable only on a restricted
region of phase space.  That is, \SubAdd\ is guaranteed to be satisfied
only if the $x_i$ and the $\tilde{x}_i$ lie within the specified region.
Our first goal, which we pursue in section~\ref{SUGRAThermo}, is to extend
the stability analysis of \cite{gspin} from one independent angular
momentum to three, and to the case of near-extreme M5-branes and M2-branes
with two and four angular momenta, respectively.  (The maximum number of
independent angular momenta is the rank of the rotation group perpendicular
to the brane).

We will restrict ourselves to the near-extreme limit of spinning branes.
For D3-branes, the string theory account of the microstates of the brane in
terms of strongly coupled ${\cal N}=4$ super-Yang-Mills theory is thought
to apply in this limit.  The near-extremal spinning geometries have a
near-horizon limit which is asymptotic to $AdS_5 \times S^5$, so if we
employ the philosophy of the AdS/CFT correspondence
\cite{juanAdS,gkPol,witHolOne}, we would say that the supergravity
thermodynamics translates directly into a characterization of thermal
states of the gauge theory.  Instability is more of a surprise in this
context, since statistical mechanical derivations of thermodynamics as a
rule imply subadditivity of the entropy.  The difficulty of treating the
strongly coupled ${\cal N}=4$ gauge theory at finite temperature prevents
us from deriving or disproving the instability from first principles.  In
section~\ref{FTanal} we do however extend the simplified field theory
treatment introduced in \cite{gspin} to the case of several angular
momenta, and we show that the results are qualitatively similar to
the supergravity predictions.

In section~\ref{CritExp} we discuss the supergravity predictions for
critical exponents at the boundary of stability.  The maximal number of
fractional critical exponents at a given point on the boundary is the
number of non-zero angular momenta.  We show in section~\ref{FTanal} that
similar critical exponents arise from the naive regulated field theory
treatment for the D3-brane.  Our analysis is not detailed on this point.

In section~\ref{EuclideanBranes} we analyze the supergravity solutions
Wick-rotated to Euclidean time.  We observe that the angular momentum
amounts in field theory to a twist imposed on the $R$-charged fields in the
field theory: the field theory partition function is altered from $\tr
e^{-\beta H}$ to $\tr R e^{-\beta H}$ where $R$ is an element of the
$R$-symmetry group.  It describes the angle through which the brane spins
during one period of the Euclidean time.  In the limit discussed in
\cite{cort} where angular momentum becomes large with energy held fixed, $R
\to (-1)^F$, so we wind up with almost supersymmetric boundary conditions.
For M5-branes compactified down to four dimensions, this leads to ${\cal
N}=4$ super-Yang-Mills theory with supersymmetry slightly broken by
boundary conditions in the Kaluza-Klein reduction.  We review how the
near-supersymmetry affects the proposals to use spinning branes as
holographic approaches to QCD, and we note a puzzle regarding the energy
scale of supersymmetry breaking and confinement.

\section{Thermodynamics of spinning branes}

\subsection{Supergravity analysis}
\label{SUGRAThermo}

The supergravity solutions for spinning branes were constructed in
\cite{CvYInt}.\footnote{Unfortunately, the   spinning M-brane solutions
 quoted in \cite{CvYInt} 
have a number  of typographical  errors.  However, these general spinning  M- as well as
D-branes    can be obtained  in a straightforward manner  from
 \cite{CvYOne}.   There
 the  general $D$-dimensional 
rotating two-charge black hole solutions (of toroidally compactified string
theory) were  given. 
The corresponding spinning  (D3, M5, M2)-branes are 
obtained by turning off one of  the two  charges of  the ($D=7,6,9$-dimensional) black
hole, and then 
lifting the solution in a straightforward way to ten and eleven dimensions,
respectively.}
  The metric for the general spinning  D3-brane  and M5-brane can also be
found in  \cite{klt,rs} and \cite{crst}, respectively.  We are interested
 primarily in the thermodynamics,
and this follows directly from the general formulas of \cite{CvYOne}.

Some parameters that enter into the thermodynamics of D3-, M2-, and
M5-branes are shown in the following table:
  \eqn{BraneProperties}{\seqalign{\span\TT \quad & \span\TC \qquad & 
    \span\TC \qquad & \span\TC \qquad & \span\TC \qquad & \span\TC}{
   brane & p & D & T_p & \alpha_p & c_p \cr
    \noalign{\vskip1\jot}\hline\noalign{\vskip1\jot}
      D3: & 3 & 7 & {\sqrt\pi \over \kappa} & 
       2 & {1 \over 8\pi^2}  \cr
      M5: & 5 & 6 & \left( {\pi \over 2\kappa^4} \right)^{1/3} &
       3 & {1 \over 96 \pi^3 } \cr
       M2: & 2 & 9 & \left( {2\pi^2 \over \kappa^2} \right)^{1/3} &
       \tf{3}{2} & {1 \over 3\sqrt{2} \pi} 
      }}
 In \BraneProperties, $\kappa$ is the ten-dimensional gravitational
coupling for the D3-brane, and the eleven-dimensional gravitational
coupling for the M2-brane and the M5-brane.  Using formulas (14) and
(15) of \cite{CvYOne}, one finds the following expressions for the
mass, charge, and spins of the branes:
  \eqn{MCSOne}{\eqalign{
   M &= {\Omega_{D-2} r_0^{D-3} \over 16\pi G_{D-2}} (D-3) \left(
    \cosh^2\alpha + {1 \over D-3} \right)  \cr
   Q &= {\Omega_{D-2} r_0^{D-3} \over 16\pi G_{D-2}} (D-3) 
     \cosh\alpha\sinh\alpha = N T_p  \cr
   J_i &= {\Omega_{D-2} r_0^{D-2} \over 16\pi G_{D-2}} 2y_i 
     \cosh\alpha  \cr
   S &= {\Omega_{D-2} r_0^{D-2} \over 16\pi G_{D-2}} 4\pi y_H 
     \cosh\alpha \ .
  }}
 Our variables are defined in terms of those of \cite{CvYOne} by 
 \eqn{Def}{
   r_0^{D-3} = 2m \qquad y_i = {\ell_i \over r_0} \qquad
    y_H = {r_H \over r_0} \ . }
 The location of the horizon, $y_H$, is the largest positive root of a
cubic or quartic equation (obtained from (16) of \cite{CvYOne}):
  \eqn{YHDef}{\seqalign{\span\TT \qquad & \span\TR}{
   D3: & (y^2+y_1^2)(y^2+y_2^2)(y^2+y_3^2) - y^2 = 0  \cr
   M5: & (y^2+y_1^2)(y^2+y_2^2) - y = 0 \cr
   M2: & (y^2+y_1^2)(y^2+y_2^2)(y^2+y_3^2)(y^2+y_4^2) - y^2 = 0 }}
 It is useful to introduce one more combination of the parameters:
  \eqn{RDef}{
   R^{D-3} \equiv r_0^{D-3} \cosh\alpha\sinh\alpha \ .
  }
 The constants $\alpha_p$ and $c_p$ were defined so that
  \eqn{acPurpose}{
   {\Omega_{D-2} \over N^{\alpha_p} 2\kappa^2} = \left\{\eqalign{
     &{c_3 \over R^8} \qquad \hbox{for D3}  \cr
     &{c_p \over R^9} \qquad \hbox{for M2 and M5.}} \right.
  }
 We can now define $E = M-Q$ and rewrite
  \eqn{MCSTwo}{\eqalign{
   e = {E \over N^{\alpha_p} V} &= {c_p \over R^{p+1}} 
    \left( {r_0 \over R} \right)^{D-3} 
    (D-3) \left( \cosh^2\alpha - \cosh\alpha\sinh\alpha + 
     {1 \over D-3} \right)  \cr
     &\approx {c_p \over R^{p+1}} {D-1 \over 2} 
     \left( {r_0 \over R} \right)^{D-3}  \cr
   j_i = {J_i \over N^{\alpha_p} V} &= {c_p \over R^p}
     2 y_i \cosh\alpha \left( {r_0 \over R} \right)^{D-2} \approx
     {c_p \over R^p} 2 y_i \left( {r_0 \over R} \right)^{(D-1)/2}  \cr
   s = {S \over N^{\alpha_p} V} &= {c_p \over R^p} 
     4\pi y_H \cosh\alpha \left( {r_0 \over R} \right)^{D-2} \approx
     {c_p \over R^p} 4\pi y_H \left( {r_0 \over R} \right)^{(D-1)/2} \ .
  }}
 The approximate expressions are for the near-extreme limit,
$\alpha\to\infty$ with $R$ held fixed.  Scaling out the power
$N^{\alpha_p} V$ from $E$, $J_i$, and $S$ allows us to write all
subsequent expression in terms of $e$, $j_i$, and $s$ alone.  That all
powers of $V$ can be absorbed in this way is dictated by the absence
of any scale in the theory; that all powers of $N$ can also be
absorbed seems to be telling us that the effective number of degrees
of freedom is on the order $N^{\alpha_p}$, as first conjectured in
\cite{ktOne}.  We can form dimensionless ratios
  \eqn{ThreeRatios}{\eqalign{
   {s^{p+1} \over e^p} &= c_p {2^{2p+1} (2\pi)^{p+1} \over (D-1)^p} y_H^{p+1}  \cr
   {j_i^{p+1} \over e^p} &= c_p {2^{2p+1} \over (D-1)^p} y_i^{p+1}  \cr
   {j_i \over s} &= {1 \over 2\pi} {y_i \over y_H} \ .
  }}

\subsubsection{Grand-canonical ensemble}

Now we wish to ask in what region of phase space $s$ is a sub-additive
function of the energy density $e$ and the angular momentum densities
$j_i$.  We regard the $j_i$ as thermodynamic variables rather than fixed
parameters, and the requirement of sub-additivity means that Legendre
transforms with respect to the $j_i$ as well as with respect to $e$ are
invertible.  Thus, in the sense described in section~III~A of \cite{CvGu1},
we are investigating stability in the grand canonical ensemble.  Stability
in the canonical ensemble, where only the Legendre transform with respect
to $e$ is required to be invertible, amounts to positivity of the specific
heat, with the $j_i$ regarded as fixed parameters.  We will treat stability
in the canonical ensemble in section~\ref{CanEns}.  We will exhibit
evidence in section~\ref{PhaseMix} that the grand-canonical ensemble is the
relevant one when the world-volume is large.  All of our actual
calculations will be done in the microcanonical ensemble, that is writing
$s = s(e,j_i)$.

The function $y_H = y_H(y_i)$ is complicated to express in closed form,
since it is the solution of a cubic or a quartic equation, cf.~\YHDef.
Using \ThreeRatios, we can write
  \eqn{SExplicit}{
   s = 2\pi\gamma e^{p/(p+1)} 
    y_H\left({j_i \over \gamma e^{p/(p+1)}}\right)  \ ,
  }
 where $\gamma^{p+1} = c_p 2^{2p+1}/(D-1)^p$, and the arguments $y_i$
of $y_H$ have been written out explicitly in terms of the $j_i$.
Fortunately, there is a more convenient way to judge the subadditivity
of \SExplicit\ than brute force.  For instance, one can require that
$e$ is concave up as a function of $s$ and the $j_i$.  Or, most
usefully, one can show that the domain of stability is the maximal
region including the points where all $j_i=0$ where the temperature $T
= \partial e/\partial s$ and the angular velocities $\Omega_i =
\partial e/\partial j_i$ are one-to-one functions of $s$ and the
$j_i$.  This is the criterion we will actually use.

By differentiating the explicit expression \SExplicit\ with respect to
$j_i$, one can show that 
  \eqn{OmegaTRatio}{
   {\Omega_i \over T} = -2\pi {\partial y_H \over \partial y_i} \ .
  }
 It is useful to define the dimensionless ratios
  \eqn{XDef}{
   x_i = {y_i \over y_H} = 2\pi {j_i \over s} \ .
  }
 Because of the scaling which follows from conformal invariance, any
dimensionless thermodynamic quantity (like $e^p/s^{p+1}$) can be
expressed in terms of the $x_i$.  This is true in particular of
$\Omega_i/T$: we have
  \eqn{pyFunct}{
   {\partial y_H \over \partial y_i} = 
    {x_i \over 1+x_i^2} W  \qquad\hbox{where}\quad
    W = {2 \over 3-D+2\sum_{j=1}^{\left\lfloor {D-1 \over 2} \right\rfloor} {x_j^2 \over 1+x_j^2}} \ .
  }
 The sum over $j$ in $W$ is over all the independent angular momenta,
of which there are $\lfloor (D-1)/2 \rfloor$.  The most
straightforward way to prove \pyFunct\ is to use the relations
  \eqn{DiffRel}{\eqalign{
   y_H^{3-D} &= \prod_i (1+x_i^2)  \cr
   (3-D) {dy_H \over y_H} &= \sum_i {2x_i \over 1+x_i^2} dx_i  \cr
   {dx_i \over x_i} &= {dy_i \over y_i} - {dy_H \over y_H} \ .
  }}
 The first of these is a rephrasing of \YHDef; the second follows from
the first by logarithmic differentiation; and the third follows from
\XDef\ also by logarithmic differentiation.  Again because of the
various scaling laws, the relations $T = \partial e/\partial s$ and
$\Omega_i = \partial e/\partial j_i$ are one-to-one if and only if
\pyFunct\ can be inverted to express the $x_i$ in terms of the
$\partial y_H/\partial y_i$.  At the boundary of a region of
$x_i$-space on which \pyFunct\ can be inverted, at least one of the
eigenvalues of the matrix $\left[ {\partial \over \partial x_i}
{\partial y_H \over \partial y_j} \right]$ goes to zero or infinity.
Provided we can show that $W$ is finite on the entire region of
stability, its boundary will be given by the solution of $\det \left[
{\partial \over \partial x_i} {\partial y_H \over \partial y_j}
\right] = 0$ closest to the origin, $x_i=0$.  Extracting an overall
factor of $W^{\lfloor (D-1)/2 \rfloor} \prod_k {1 \over 1+x_k^2}$, the
determinant equation becomes
  \eqn{DetRel}{
   D(x_i) \equiv \det\left[ x_i {\partial\log W \over \partial x_j} + 
    \delta_{ij} {1-x_i^2 \over 1+x_i^2} \right] = 0 \ .
  }
 This determinant has a simple closed form:
  \eqn{SolvedD}{
   D(x_i) = {(D-3) P(-1) + 2 P'(-1) \over (D-3) P(1) - 2 P'(1)}
  }
 where $P(\lambda)$ is defined by
  \eqn{PDef}{
   P(\lambda) = \prod_k (1 + \lambda x_k^2)
    = 1+\sum_{\ell =1}^{\left\lfloor {D-1 \over 2} \right\rfloor} \lambda^\ell 
    \sum_{i_1>i_2>\ldots>i_\ell} 
        \prod_{j=1}^\ell x_{i_j}^2 \ .
  }
 Note that $D(x_i) = 1$ when all the $x_i$ vanish.  Thus the region of
stability is the domain including $x_i=0$ where $D(x_i) \geq 0$.  If
all but one angular momentum is zero, then
  \eqn{OneJ}{
   D(x) = {(D-3) - (D-5) x^2 \over (D-3) + (D-5) x^2}
  }
 Which indicates that the region of stability for a single angular momentum
is $x^2 \leq {D-3 \over D-5}$.  Apparently by coincidence, $\alpha_p = {D-3
\over D-5}$ as well.  The bound $x^2 \leq \alpha_p$ is in agreement with
equation (17) of \cite{gspin} in the case $p=3$, and with the bounds
derived in \cite{cai} for $p=2$ and $p=5$.

It is straightforward to verify that the denominators of $W$ in
\pyFunct\ and $D(x_i)$ in \DetRel\ differ only by a factor $-P(1)$, and
that both these denominators are nonzero and finite throughout the
domain including $x_i = 0$ where the numerator of $D(x_i)$ is
positive.  Thus an equivalent condition of stability to $D(x_i) \geq
0$ is
  \eqn{PStab}{
   (D-3) P(-1) + 2 P'(-1)= 
   (D-3)+ \sum_{\ell =1}^{\left\lfloor {D-1 \over 2} \right\rfloor}(-1)^\ell (D-3-2\ell)
   \sum_{i_1>i_2\cdots>i_\ell }\prod_{j=1}^\ell x_{i_j}^2\geq 0 \ .
  }
 For the three types of branes we are considering, \PStab\ specializes
to 
  \eqn{PStabCases}{\seqalign{\span\TT \qquad & \span\TR}{
   D3: & 2 - x_1^2 - x_2^2 - x_3^2 + x_1^2 x_2^2 x_3^2 \geq 0  \cr
   M5: & 3 - x_1^2 - x_2^2 - x_1^2 x_2^2 \geq 0\cr
    M2: & 3 - 2 (x_1^2 + x_2^2 + x_3^2 + x_4^2)   \cr
       & \quad {} + (x_1^2 x_2^2 + x_1^2 x_3^2 + x_1^2 x_4^2 + 
                x_2^2 x_3^2 + x_2^2 x_4^2 + x_3^2 x_4^2)  \cr
       & \quad {} - x_1^2 x_2^2 x_3^2 x_4^2 \geq 0 \ .
   }}

 Our methods can also be used to judge the stability of rotating black
strings in five and six dimensions arising from intersections of spinning  M-branes:
$M5 \perp M5 \perp M5$ for five dimensions, and $M5 \perp M2$ for six
dimensions.  Because the world-volume is effectively two-dimensional, one
does not expect any phase transition at finite temperature, on account of
the Mermin-Wagner theorem \cite{MerminWagner}.  Indeed, it is
straightforward to verify by direct computation that the entropy function
is subadditive.  It is also possible to trace through an analysis like the
one in section~\ref{SUGRAThermo}, introducing a combination of parameters
$R_i^{D-3} = r_0^{D-3}\cosh\alpha_i\sinh\alpha_i$ for each constituent
family of branes, and verifying that \SolvedD\ is always satisfied.  (We
note however that \PStab\ is {\it not} always satisfied for the six-dimensional
case: a zero in the denominator of \SolvedD\ cancels a zero in the
numerator.)

\subsubsection{Canonical Ensemble}
\label{CanEns}

For the sake of completeness we also derive the subadditivity constraints
for the canonical ensemble. This is the case where the angular momenta
(``charges'') of rotating branes are kept fixed, so that subadditivity of
the entropy is exactly equivalent to positivity of the specific heat at
constant $j_i$.  As we will demonstrate, it may be difficult to keep the
$j_i$ uniform over the entire world-volume outside of the grand-canonical
region of stability.  However, uniform $j_i$ will be our working assumption
in this section.  It is a general result, demonstrated in section~III~A of
\cite{CvGu1} and verifiable directly from comparing \eno{canonical} to
\eno{PStabCases}, that the region of stability in the canonical ensemble is
larger than in the grand canonical ensemble.  The canonical stability
constraints for spinning branes with one angular momentum were first
studied by Cai and Soh in \cite{cai}.
 
Using \MCSTwo, \ThreeRatios, \SExplicit, \pyFunct, and \DiffRel, it is
fairly straightforward to reduce the inequality $\partial^2 s/\partial e^2
\leq 0$ to
 \begin{eqnarray}
{\rm D3:}\  & &
2
+5\sum_{i=1}^3 x_i^2
-\sum_{i=1}^3 x_i^4
+10\sum_{i<j}^3x_i^2 x_j^2
+16x_1^2 x_2^2 x_3^2
+2\sum_{\{i<j\}\ne k}^3x_i^2x_j^2 x_k^4 \nonumber\\
& &
+3\sum_{\{i<j\} \ne k}^3x_i^4 x_j^4 x_k^2
+5x_1^4x_2^4x_3^4
\ge 0 \ , \nonumber\\
{\rm M5:} \ & &
9
+36(x_1^2+x_2^2)
-9(x_1^4+x_2^4)
+76x_1^2x_2^2
+8(x_1^4x_2^2+x_2^4x_1^2)
+21x_1^4x_2^4
\ge 0 \ ,\nonumber\\
{\rm M2:}\ & &
 9
+18\sum_{i=1}^4x_i^2
+34\sum_{i<j}^4x_i^2x_j^2
+60\sum_{i<j<k}^4x_i^2 x_j^2 x_k^2
+2\sum_{i\ne j}^4x_i^2x_j^4
-3\sum_{i<j}^4x_i^4x_j^4
+6\sum_{\{i<j\}\ne  k}^4 x_i^2x_j^2x_k^4\nonumber\\
 &&
 +96x_1^2x_2^2x_3^2x_4^2\
-2\sum_{\{i<j\}\ne  k}^4 x_i^4x_j^4x_k^2
+12\sum_{\{i<j<k\}\ne \ell}^4 x_i^2x_j^2x_k^2x_{\ell}^4
\nonumber\\
& &
+2\sum_{\{i<j\}\ne  \{k<\ell\}}^4 x_i^4x_j^4x_k^2x_{\ell}^2
+6\sum_{\{i<j<k\}\ne\ell}^4 x_i^4x_j^4x_k^4x_{\ell}^2
+9x_1^4x_2^4x_3^4x_4^4\ge 0\ , \\
\label{canonical}
\end{eqnarray}
where the $x_i$ are defined in \XDef.

A special case  a single angular momentum turned on, say, $x_1=x\ne 0$, 
these stability constraints
reduce to:
\begin{eqnarray}
{\rm D3}: & &\  2 +5x^2-x^4\ge 0\ ,  \nonumber \\
{\rm M5}: & &\  1+4x^2- x^4\ge 0\ ,\nonumber \\
{\rm M2}: & &\  1+2x^2\ge 0\ .\,  \label{canon1}
\end{eqnarray}
 which are identical to the constraints derived in \cite{cai}. Note that
the M2-brane with a single angular momentum is always stable, but that with
two angular momenta or more an instability is possible. 
(For example,  when one approaches  criticality along a
 special  directions of 
   $n$ equal  non-zero angular momenta
    the critical values become  $ x_c \approx 2.118431$  and 
$x_c \approx 2.193567$ for  
   $n=2$  and $n=3$, respectively.)

\subsection{Critical Exponents}
\label{CritExp}

We will focus on the grand canonical ensemble.  Critical exponents
characterize the behavior near a critical point of the grand potential,
also called the Gibbs potential, and defined as $\Xi=E-ST-\sum_i\Omega_i
J_i$.  Conformal invariance implies $\Xi=-\textstyle{1\over p}E$.  Using
\MCSTwo\ and \ThreeRatios, one derives
 \eqn{GrandPot}{\eqalign{
 \xi\equiv {\Xi\over{VN^{\alpha_p}}}= &-{p^p\over{(p+1)^{p+1}}}
 T^{p+1}{s^{p+1}\over e^p}
 \left(1+{j_i\over s}{\Omega_i\over T}\right)^{p+1}\cr
 =&-CT^{p+1}\prod_{i}(1+x_i^2)^{{p+1}\over {3-D}}\left({{3-D}\over 
 {3-D+2\sum_j
 {{x_j^2}\over {1+x_j^2}}}}\right)^{p+1}\ , 
 }}
 where $C={p^p\over {(p+1)^{p+1}}}(2\pi\gamma)^{p+1}$.  One can also show
that the inverse temperature of the solution is
\eqn{Beta}{
 \beta\equiv {1\over T}= {{8\pi p}\over{(D-1)(p+1)}}R 
 \left({R\over r_0}\right)^{{D-3}\over
 {p+1}}
 \prod_{i}(1+x_i^2)^{1\over {3-D}}\left({{3-D}\over 
 {3-D+2\sum_j
 {{x_j^2}\over {1+x_j^2}}}}\right)\ .}

The expression \GrandPot\ for $\xi$ is in terms of the $x_i$, and it is
straightforward to make a Taylor expansion around a point $x_{ic}$ on the
boundary of stability.  There is no singular behavior in this expansion.
However, if $\xi$ is expanded around the chosen point $x_{ic}$ in powers of
$\Omega_i-\Omega_{ci}$, there are in general fractional critical exponents,
$\gamma_{ci}$.  This is possible when in the expansion 
 \begin{equation}\eqalign{
\omega_i-\omega_{ic}&= 2\pi\left(A_{ij}\Bigg|_{x_k=x_{kc}}(x_j-x_{jc})+
{{\partial A_{ij}}\over {\partial
x_k}}\Bigg|_{x_l=x_{lc}}(x_j-x_{jc})(x_k-x_{kc})+\cdots \right)  \cr
  &\hbox{where}\quad A_{ij} = {\partial \over \partial x_i} 
   {\partial y_H \over \partial y_i} \ ,}
\label{expan}\end{equation}
 the first term on the right hand side vanishes.  There will always be
directions of approach in $x_i$-space where this term does vanish: $A_{ij}$
has vanishing determinant at the boundary of stability, and if we set $x_i
= x_{ic} - \eta v_i$ where $A_{ij} v_j = 0$, then
  \begin{eqnarray}
   \omega_i - \omega_{ic} &\sim& c_2 \eta^2 + c_3 \eta^3 + 
     \ldots \label{omegaLim}  \\
   \xi - \xi_c &\sim& \tilde{c}_2 \eta^2 + \tilde{c}_3 \eta^3 + 
     \ldots \label{xiLim} \ ,
  \end{eqnarray}
 for some coefficients $c_i$ and $\tilde{c}_i$.  There is no linear term in
$\eta$ in \eno{omegaLim} by construction of $v_i$.  There is also no linear
term in \eno{xiLim} because such a term would cause some component of the
angular momentum to become infinite as $\eta \to 0$, and this we can rule
out with calculations in the microcanonical ensemble.  Eliminating $\eta$
one can obtain an expansion for $\xi$ of the form
  \eqn{xiExpandForm}{
   \xi \sim \hbox{(analytic)} + |\omega_i - \omega_{ic}|^{2-\gamma} +
     \ldots \ ,
  }
 where the term shown explicitly is the leading non-analytic term.  In the
absence of numerical coincidences such as the $\eta^2$ term vanishing in
\eno{omegaLim}, or the relative coefficients between $\eta^2$ and $\eta^3$
being identical in \eno{omegaLim} and \eno{xiLim}, we will have $\gamma =
1/2$.  It is important to keep in mind that the exponent $\gamma$ is a
function both of the particular point on the boundary of stability one is
approaching and of the direction of approach.  In the cases we have
examined, generically $\gamma = 1/2$ with some isolated exceptions which we
will now describe.

Let us take set $n$ angular momenta equal, so that $x_1 = \ldots = x_n =
x$, and send $x$ to its critical value $x_c$ with the other $x_i$
vanishing.  In this case the expression for the common value $\omega$ of
$\omega_i\equiv{\Omega_i\over T}$ in terms of $x$ can be inverted
explicitly, and it takes the form:
 \eqn{xiom}{x=\textstyle{1\over{2n+3-D}}\left(
-\textstyle{{2\pi}\over {\omega}}\pm [\left(\textstyle{{2\pi}\over {\omega
}}\right)^2+(2n+3-D)(D-3)]^{1/2}\right), \ \
}
 Half integer powers appear in the expansion of $x$ around
$\Omega-\Omega_c$ when $(2n+3-D)(D-3)<0$, and this leads to $\gamma = 1/2$.
This situation obtains when $n=1$ for all $D$, and also for $n=2$ in the
case of the M2-brane.  In fact $n=1$ is essentially the generic case:
approach to the boundary along a normal direction away from any special
point.  When $n>1$ for the D3-brane or the M5-brane, and when $n>2$ for the
M2-brane, \eno{xiom} leads to an expansion of $\xi$ which involves only
integer powers of $\omega - \omega_c$.

 
On account of \eno{expan}, there can be as many independent directions where
$\gamma$ is fractional as there are independent vectors annihilated by
$A_{ij}$.  In order to determine when $A_{ij}$ in (\ref{expan}) has
multiple zero eigenvalues, an efficient method is to determine the entire
locus of points where the polynomials in \PStabCases\ vanish.  Let us call
one of these polynomials $p(x_i)$, and note that $p = \lambda_1 \ldots
\lambda_n$ up to a factor which is never $0$, where $\lambda_i$ are the
eigenvalues of $A_{ij}|_{x_k=x_{kc}}$.  The locus $p=0$ is the union of the
vanishing sets of the $\lambda_i$.  At a generic point on $p=0$, only one
$\lambda_i = 0$.  Near a point of $p=0$ where the vanishing set of one
eigenvalue crosses the vanishing set of another, the equation $p=0$ does
not define a manifold; rather, it defines two or more manifolds crossing.
We can find these points by solving $p = 0$ and $dp \equiv
\sum_{i=1}^{\left\lfloor {D-1 \over 2} \right\rfloor} {{\partial p}\over
{\partial x_i}}dx_i = 0$ simultaneously.  For the M5-brane, there is no
solution: $f = 0$ truly defines a manifold, and the two eigenvalues of
$A_{ij}$ never simultaneously vanish.  For the D3-brane and the M2-brane,
the only solutions are where all $x_i^2 = 1$.  For the D3-brane, this is a
branch point where two sheets of $p=0$ cross, and correspondingly two
eigenvalues of $A_{ij}$ vanish.  For the M2-brane, three sheets of $f=0$
cross, so three eigenvalues of $A_{ij}$ vanish.  In general, the number of
eigenvalues vanishing at a branch point is indicated by the lowest degree
term in the Taylor expansion of $p$ around the branch point: that is, the
order of the branch point.

\subsection{Field theory analysis}
\label{FTanal}

In \cite{gspin} a field theory analysis of spinning D3-branes was
attempted, using the fact that transverse angular momentum on the brane
amounted to R-charge.  In this section we will extend the analysis from one
angular momentum to three.  We will also briefly consider the M2-brane and
the M5-brane.

The on-shell degrees of freedom in a single D3-brane's world-volume theory
are two vector boson states (counting positive and negative helicity
separately), six scalar states, and eight fermion states.  The R-charges of
these states are specified by vectors in the weight lattice of $SO(6)$.
The on-shell degrees of freedom for a single M2-brane are eight scalar
states and eight fermion states, with R-charges specified by weight vectors
of $SO(8)$.  The on-shell degrees of freedom for a single M5-brane are
three self-dual two-form states, five scalar states, and eight fermion
states, with R-charges specified by weight vectors of $SO(5)$.  All these
states are tabulated in equation \eno{WeightVectors}:
  $$\eqalign{
    \seqalign{\span\TT}{on-shell D3-brane states, \cr\noalign{\vskip-6\jot}
      $SO(6)$ quantum numbers} &\qquad\left\{
        \seqalign{\span\TT\quad & \span\TR}{
          vector: & 2(0,0,0)  \cr\noalign{\vskip-5\jot}
          scalar: & (\pm1,0,0) + (0,\pm1,0) + (0,0,\pm1)  
            \cr\noalign{\vskip-5\jot}
          fermion: & \left( \pm\tf{1}{2},\pm\tf{1}{2},\pm\tf{1}{2} \right)} 
    \right. }$$\eqn{WeightVectors}{\eqalign{
    \seqalign{\span\TT}{on-shell M2-brane states, \cr\noalign{\vskip-6\jot}
      $SO(8)$ quantum numbers} &\qquad\left\{
        \seqalign{\span\TT\quad & \span\TR}{
          scalar: & (\pm1,0,0,0) + (0,\pm1,0,0) \cr\noalign{\vskip-5\jot}
           &\quad + (0,0,\pm1,0) + (0,0,0,\pm1)  \cr\noalign{\vskip-5\jot}
          fermion: & \left( \pm\tf{1}{2},\pm\tf{1}{2},\pm\tf{1}{2},
           \pm\tf{1}{2} \right) \ ,  \cr\noalign{\vskip-5\jot}
           &\quad\hbox{even number of -}}
    \right.  \cr
    \seqalign{\span\TT}{on-shell M5-brane states, \cr\noalign{\vskip-6\jot}
      $SO(5)$ quantum numbers} &\qquad\left\{
        \seqalign{\span\TT\quad & \span\TR}{
          two-form: & 3(0,0)  \cr\noalign{\vskip-5\jot}
          scalar: & (\pm1,0) + (0,\pm1) + (0,0)  \cr\noalign{\vskip-5\jot}
          fermion: & 2 \left( \pm\tf{1}{2},\pm\tf{1}{2} \right)} 
    \right.
  }}
 For $p = 2$, $3$, or $5$, let $\vec{\alpha}$ be a vector which runs over the
sixteen $p$-brane weights given above, and set $s_{\vec{\alpha}} = 1$ when the
corresponding particle is a boson and $s_{\vec{\alpha}} = -1$ when it is a fermion.
We will calculate in the grand canonical ensemble with ``voltages''
$\Omega_1$, $\Omega_2$, \ldots, $\Omega_{\lfloor (D-1)/2 \rfloor}$
corresponding to the elements of the Cartan subalgebra of the R-symmetry
group (for instance, in the case of D3-branes, we have $\Omega_1$,
$\Omega_2$, and $\Omega_3$ for the three elements of the Cartan subalgebra
of $SO(6)$).  Denoting $\vec{\omega} = (\omega_1,\omega_2,\ldots,\omega_{\lfloor
(D-1)/2 \rfloor})$ where $\omega_i = \Omega_i/T$, we have in analogy to
(26) of \cite{gspin} an expression
  \eqn{XiInt}{
   \xi = T \int {d^p k \over (2\pi)^p} \sum_{\vec{\alpha}} 
     s_{\vec{\alpha}} \log(1-s_{\vec{\alpha}} e^{-\beta |\vec{k}| + \vec{\alpha} \cdot \vec{\omega}}) = 
    -{T^{p+1}\Gamma(p) \over 2^{p-1} \pi^{p/2} \Gamma(p/2)} \sum_{\vec{\alpha}} s_{\vec{\alpha}} 
      \Li_{p+1}(s_{\vec{\alpha}} e^{\vec{\alpha} \cdot \vec{\omega}}) 
  }
 for the grand potential per unit volume, $\xi$, which is also the
pressure.  As in the case of only one angular momentum, the integrals
diverge for any real nonzero $\vec{\omega}$ because there are charged
massless bosons.  The integrals in \XiInt\ are all special cases of 
  \eqn{LiIntTwo}{
   \int_0^\infty dk \, k^{n-1} \log(1 - e^{\mu-k})
    = -\Gamma(n) \Li_{n+1}(e^\mu) \ ,
  }
 which is carried out assuming complex $\mu$ and using analytic
continuation.  The difficulty in defining the integrals for real $\vec{\omega}$
is apparent from the branch cuts across the positive real $\mu$ axis of
$\Li_{p+1}(e^\mu)$.  In \cite{gspin} it was proposed to define the
integrals using the principle value prescription, which in this case
amounted to replacing the polylogarithms by their real parts.  This
prescription together with the identity 
  \eqn{LiIdentity}{
   \Li_n(e^\mu) + (-1)^n \Li_n(e^{-\mu}) = 
    \sum_{j=0}^{\lfloor n/2 \rfloor} {2 \zeta(2j) \over
     (n-2j)!} \mu^{n-2j} \pm {i\pi \over (n-1)!} \mu^{n-1} \ .
  }
 leads to
  \eqn{XiPoly}{
   -{\xi \over T^4} \equiv g(\vec{\omega}) = 
     {\pi^2 \over 6} + {1 \over 4} \sum_i \omega_i^2 + 
     {1 \over 32\pi^2} \left( \sum_i \omega_i^2 \right)^2 - 
     {1 \over 16\pi^2} \sum_i \omega_i^4 
  }
 for the D3-brane, and to 
  \eqn{XiPolyMFive}{\eqalign{
   -{\xi \over T^6} &= {\pi^3 \over 30} + 
     {\pi \over 24} (\omega_1^2 + \omega_2^2) + 
     {1 \over 96\pi} (\omega_1^2 + \omega_2^2)^2 +
     {1 \over 48\pi} (\omega_1^4 + \omega_2^4)  \cr
    &\qquad{} + {1 \over 1152\pi^3} (\omega_1^2 + \omega_2^2)^3 - 
     {1 \over 288\pi^3} (\omega_1^6 + \omega_2^6)
  }}
 for the M5-brane.  For the M2-brane, the grand potential is
transcendental, and cannot be simplified substantially beyond \XiInt.  At
small $\omega_i$ one has the expansion
  \eqn{XiPolyMTwo}{
   -{\xi \over T^3} = {7 \zeta(3) \over \pi} - 
     \sum_i {\omega_i^2 \over 4\pi} \left( \log {\omega_i^2 \over 4} - 
      3 \right) + \ldots \ .
  }
 For $N$ coincident D3-branes branes in the supergravity limit, the
world-volume theory is strongly interacting.  We nevertheless suggest
\XiPoly\ as a first approximation to the grand potential of
$3+1$-dimensional ${\cal N}=4$ $SU(N)$ gauge theory at high temperature,
where now we have scaled a factor of $N^2$ out of $\xi$ as in
section~\ref{SUGRAThermo}.  The speculation is that even though the
fundamental excitations are strongly interacting, for the purposes of
thermodynamics we can think of $O(N^2)$ quasi-particle excitations which
have the same multiplet structure as the fundamental degrees of freedom.
For the M2-brane, agreement with supergravity requires $O(N^{3/2})$
quasi-particles, whereas for the M5-brane, $O(N^3)$ are required.  These
peculiar scalings are borne out by absorption calculations
\cite{gkSchwing}, but are otherwise poorly understood.

For the rest of this section we will focus on the D3-brane.

Thermodynamic stability requires that $\xi$ be a subadditive function of
$T$ and the $\Omega_i$.  Because of the scaling laws due to conformal
invariance, this amounts to being able to express the $\omega_i$ as
single-valued functions of the ratios $j_i/s$.  It is straightforward to
check that this is possible on the region of $\omega$ space containing the
origin where
  \eqn{SuperConvex}{
   \det\left( 4 g {\partial^2 g \over \partial\omega_k\partial\omega_\ell} - 
    3 {\partial g \over \partial\omega_k} 
      {\partial g \over \partial\omega_\ell} + 
    \omega_i {\partial g \over \partial\omega_k}
      {\partial^2 g \over \partial\omega_i \partial\omega_\ell} - 
    \omega_i {\partial g \over \partial\omega_i}
      {\partial^2 g \over \partial\omega_k \partial\omega_\ell}
   \right) \geq 0 \ .
  }
 In \SuperConvex, which generalizes (42) of \cite{gspin}, the index
$i$ is summed over in the last two terms, and $k$ and $\ell$ are the
indices of the $3\times3$ matrix whose determinant is taken.

The mean field theory analysis of \cite{gspin} also generalizes in a
simple way.  We will omit some details which do not depend on the
number of angular momenta.  The basic idea is that there is some
``self-field'' contribution to the voltage, which we express as
  \eqn{SelfField}{
   \varphi = \omega + f(j/T^3) \ .
  }
 Here $\varphi$ is the total voltage, $\omega$ is the applied voltage, and
$f(j/T^3)$ is the self-field contribution, given in terms of an unknown
function $f$, and in terms of an R-charge density $j$ obtained by
integrating over the thermal occupation factors (using the principle value
prescription) in the presence of the total voltage $\varphi$.  For
notational simplicity we have dropped the arrows from $\varphi$, $\omega$,
and $j$, but they are indeed three-dimensional vectors, and derivatives
with respect to them are also three-dimensional vectors.  Thus the
permeability $\mu = \partial\varphi/\partial\omega$ is a $3\times3$ matrix
which can depend on $\omega$.  The simplest extension of \cite{gspin} is to
assume $f(x) = -x$, which implies that the permeability at zero voltage is
$\mu = \tf{2}{3} {\bf 1}$.  That is, the D3-brane responds diamagnetically
to an applied voltage.  We will show how this ansatz reproduces the
critical behavior found in supergravity.

Mean field theory leads to a parametric expression for the free
energy:
  \eqn{XiParametric}{\eqalign{
   -{\xi \over T^4} &\equiv \tilde{g}(\omega) = g(\varphi)  \cr
   \omega &= \varphi + {\partial g \over \partial\varphi} \ .
  }}
 The first line in \XiParametric\ uses \XiInt\ with $\omega$ replaced
by $\varphi$.  The second line is a rephrasing of \SelfField\ using
the ansatz $f(x) = -x$.  The function $g(\varphi)$ is the same as
appeared in \XiPoly.  Since $g(\varphi)$ is analytic, the only way
nonanalyticity can arise is through inverting the relation between
$\omega$ and $\varphi$.  Nonanalyticity can arise in this context only
if the inverse permeability, 
  \eqn{InversePermeability}{
    \mu^{-1} = {\partial\omega \over \partial\varphi}
      = {\bf 1} + {\partial^2 g \over \partial\varphi^2} \ ,
  }
 acquires a zero eigenvalue.  Thus the boundary of stability is
determined by $\det \mu^{-1} = 0$, which is equivalent to 
  \eqn{LongForm}{\eqalign{
   & 576 \pi^6 - 48 \pi^4 (\varphi_1^2 + \varphi_2^2 + \varphi_3^2) - 
    20 \pi^2 (\varphi_1^4 + \varphi_2^4 + \varphi_3^4) \cr & +
    8 \pi^2 (\varphi_1^2 \varphi_2^2 + \varphi_1^2 \varphi_3^2 +
     \varphi_2^2 \varphi_3^2) - 
    (\varphi_1^6 + \varphi_2^6 + \varphi_3^6) + 
    6 \varphi_1^2 \varphi_2^2 \varphi_3^2 \cr & + 
    \varphi_1^2 \varphi_2^4 + \varphi_1^2 \varphi_3^4 + 
    \varphi_2^2 \varphi_1^4 + \varphi_2^2 \varphi_3^4 + 
    \varphi_3^2 \varphi_1^4 + \varphi_3^2 \varphi_2^4 \geq 0 \ .
  }}
 This complicated expression reduces to $|\varphi_1| \leq 2\pi$ when
$\varphi_2 = \varphi_3 = 0$.  It exhibits the same qualitative behavior
as \PStabCases, in that the boundary of stability is a generically a
single zero of the left hand side, with the sole exception of all
$\varphi_i^2$ equal, when the boundary of stability becomes a double
zero.  As before, this indicates that generically only one eigenvalue
crosses through $0$ at the boundary, but for all angular momenta equal
two eigenvalues vanish at the same time.  We obtain the same critical
behavior from the mean field theory analysis as from the supergravity,
and for basically the same reasons: making the expansions
  \eqn{MFTExp}{\eqalign{
   \omega-\omega_c &= ({\bf 1} + g''(\varphi_c)) (\varphi-\varphi_c) + 
    \tf{1}{2} g'''(\varphi_c) (\varphi-\varphi_c)^2 + \ldots \cr
   -{\xi \over T^4} &= g(\varphi_c) + g'(\varphi_c)(\varphi-\varphi_c) + 
    \tf{1}{2} g''(\varphi_c) (\varphi-\varphi_c)^2 + \ldots \ ,
  }}
 we see that $\gamma = 1/2$ for variations of $\omega$ in the directions in
which $\mu^{-1}$ has a zero eigenvalue, while $\gamma = 0$ in the other
directions.  The algebra is the same as in~\eno{omegaLim} and~\eno{xiLim},
except that there $\eta$ represented a deviation parametrized in terms of
$x-x_c$, whereas in~\MFTExp\ it would be a deviation parametrized in terms
of $\varphi-\varphi_c$.

\section{Phase mixing}
\label{PhaseMix}

So far our goal has been to demonstrate that spinning branes with a
sufficiently high angular momentum become thermodynamically unstable.  We
have not answered the obvious question: what form does the instability
take?  In this section we will entertain two possibilities: 1) the branes
fragment and fly apart in the transverse dimensions; 
and 2) the angular momentum density
localizes on the branes.  For the sake of specificity, we will focus
exclusively on $N$ coincident D3-branes with only one angular momentum.  In
the notation of \cite{gspin},
  \eqn{ChiDef}{
   \chi = {27 \pi^2 \over 8N^2} {(J/V)^4 \over (E/V)^3} > {1 \over 3}
  }
 puts us outside the region of stability as calculated in the grand
canonical ensemble.

The naive expectation is that D3-branes with $\chi > 1/3$ vigorously shed
their angular momentum by radiating either closed strings or individual
D3-branes.  In an examination of the radial potentials for closed string
emission and for equatorial motion of a test D3-brane in the background of
$N$ spinning D3-branes, we found no evidence for a special unstable mode
developing at $\chi = 1/3$.  This is not necessarily a problem: as
commented upon in the introduction, the nature of the instability is not so
much that there is some runaway mode of the system by itself, but rather
that it is impossible to establish equilibrium between it and a heat bath.

The following simple estimate suggests that for $\chi > 1/8$ D-brane
fragmentation is thermodynamically allowed.\footnote{We thank F.~Larsen,
E.~Martinec, and P.~Kraus for a stimulating discussion in which this
calculation was suggested to us.}  Suppose the the $N$ D3-branes split into
$k$ equal fragments.  As the fragments fly apart, they can acquire orbital
angular momentum.  As an idealization of this complicated dynamical fission
process, let us suppose that most of the angular momentum winds up in the
orbital motion of the fragments, and that most of the energy winds up in
their non-extremality.\footnote{If we accounted for the energy that goes
into Hawking radiation and the motion of the fragments, the bound would be
$\chi > \chi_0$ for some $\chi_0 > 1/8$.}  As a further idealization, we
will assume that the process occurs uniformly across the D3-brane
world-volume.  By compactifying the world-volume on a small $T^3$ one could
ensure such uniformity.  From \cite{gspin} we take the formula
  \eqn{sejN}{
   s(e,j;N) = {2^{5/4} 3^{-3/4} \sqrt{\pi N} e^{3/4} \over
    \sqrt{\sqrt{1+\chi} + \sqrt{\chi}}}
  }
 for the entropy per unit world-volume in terms of the energy and angular
momentum per unit world-volume.  Since we will be considering different
values of $N$, we have reverted to the conventions of \cite{gspin}: $e =
E/V$, $j = J/V$, and $s = S/V$, without the factors of $N^2$ in the
denominator which we used in section~\ref{SUGRAThermo}.  The ratio of the
entropy of the final state ($k$ identical non-spinning fragments) to the
initial state (a single cluster with spin) is
  \eqn{sRatio}{
   {s_f \over s_i} = 
    {k s\left( {e \over k},0;{N \over k} \right) \over
     s(e,j;N)} = \sqrt{{\sqrt{1+\chi}+\sqrt{\chi} \over \sqrt{k}}} \ .
  }
 The process is entropically favored if $s_f > s_i$, which is equivalent to
$\sqrt{k} \leq \sqrt{1+\chi} + \sqrt{\chi}$.  Kinematics forbids the $k=1$
process, so we get the lowest critical $\chi$ from $k=2$ (two fragments):
$\chi > 1/8$, as advertised.  With $k=3$ (three fragments) we get $\chi >
1/3$, which coincidentally corresponds exactly to the boundary of the
domain where $s(e,j)$ is subadditive.  Making the fragments unequal only 
increases the bound.  For very large $\chi$, large values of $k$ will be
accessible, and we interpret the fragmentation as a Jeans instability of
the disk of D3-branes described in \cite{klt}.

The above estimate does not convince us that brane fragmentation is the
dominant instability for $\chi$ slightly larger than $1/3$.  We would be
more optimistic if an instability of the horizon could be identified.  In
any case, brane fragmentation would violate singularity theorems, so we may
expect it to have a very long time-scale in the classical limit of many
coincident branes.

We now turn to the second possibility: that for $\chi > 1/3$ the angular
momentum density $j$ tends to localize on the world-volume.  If they are
thermodynamically permitted, inhomogeneities in $j$ might nucleate locally
on a shorter time-scale than brane fragmentation.  In fact, the time scale
should be on the order of the inverse temperature since there is no other
scale in the problem.

Let us regard \sejN\ as an approximate expression for the entropy density
in terms of energy density $e$ and angular momentum density $j$ which are
allowed to vary over the world-volume.  Diffusion and interactions will
tend to keep the temperature $T$ and the ``voltage'' $\Omega$ (properly the
angular velocity at the black hole horizon) spatially constant.  However,
for a given $(T,\Omega)$, there are two possible values for energy density
and angular momentum density.  (That there are two and only two possible
values can be seen from the fact that the equation of state, (20) of
\cite{gspin}, is double-valued).  One of them, call it $(e,j)$, will lie in
the stable region where $s$ is subadditive; the other, call it
$(\tilde{e},\tilde{j})$, will lie in the unstable region.  Thus we are led
to consider mixed states of the two phases $(e,j)$ and
$(\tilde{e},\tilde{j})$.  We will argue that such states are entropically
favored over states with uniform $(e,j)$ for $\chi \gsim 0.325$, although
only for $\chi > 1/3$ can a uniform state evolve smoothly into such a mixed
state.  The mixed states are only saddle points of the entropy functions,
not maxima, because they involve an unstable component.

We emphasize that the following calculation is not on the same footing as
the usual analysis of phase coexistence, because one ``phase'' in the
current context is unstable.  Rather, it is intended as a preliminary probe
of how inhomogeneous $e$ and $j$ may arise near $\chi = 1/3$, based on the
idea that $T$ and $\Omega$ should remain at least approximately uniform
while these inhomogeneities are first forming.  Because we are relying on
the known supergravity solutions as our guide to the thermodynamics, we are
not equipped to address the questions of coherence length, surface tension,
and stability of phase boundaries.  Ideally, one would like to establish
the existence of a second stable phase, perhaps one where the branes are
separated in the transverse dimensions.  Then a more standard equilibrium
analysis would become possible.

Assume that a fraction $\lambda$ of the spatial volume is filled by the
unstable phase $(\tilde{e},\tilde{j})$, and the remaining fraction
$1-\lambda$ is filled by the stable phase, $(e,j)$.  Then the average
energy density, angular momentum density, and entropy density are
  \eqn{ejConstraint}{\eqalign{
   e_{av} &= \lambda \tilde{e} + (1-\lambda) e  \cr
   j_{av} &= \lambda \tilde{j} + (1-\lambda) j  \cr
   s_{av} &= \lambda s(\tilde{e},\tilde{j};N) + (1-\lambda) s(e,j;N) \ ,
  }}
 where $s(e,j;N)$ is the function \sejN.  With $e_{av}$ and $j_{av}$ fixed
we want to extremize $s_{av}$ by varying $\lambda$.  That is, in an isolated
system where the total energy and angular momentum are fixed, we extremize
the total entropy by distributing energy and angular momentum optimally
between the two phases.  This problem is mathematically well-posed because
$\tilde{e}$ and $\tilde{j}$ can be determined as functions of $e$ and $j$,
and then the first two lines of \ejConstraint\ amount to two (non-linear)
constraints on three variables, leaving $\lambda$ as the only free
variable.  

More explicitly, for given $(T,\Omega)$, there are two solutions
to the equations
  \eqn{Derivs}{\eqalign{
   \left( \partial s \over \partial j \right)_e &= 
    -{\Omega \over T} = -{\pi \sqrt{2} \chi^{1/4}
     \sqrt{\sqrt{1+\chi}-\sqrt\chi} \over \sqrt{1+\chi}}  \cr
   \left( \partial s \over \partial e \right)_j &=
    {1 \over T} = {\sqrt\pi 3^{1/4}/2^{3/4} \over e^{1/4}
     \sqrt{\sqrt{1+\chi}-\sqrt\chi} \sqrt{1+\chi}} \ ,
  }}
 and if one of them is $(e,j)$, then the other, $(\tilde{e},\tilde{j})$,
can be expressed as
  \eqn{wDef}{\eqalign{
   &\tilde{e} = e \left( {1+\chi \over 1+\tilde\chi} \right)^4
    {\tilde\chi \over \chi} \ , \quad
   \tilde{j} = j \left( {1+\chi \over 1+\tilde\chi} \right)^3
    {\tilde\chi \over \chi}  \cr
   &\hbox{where} \quad 
    w = {\sqrt\chi (\sqrt{1+\chi}-\sqrt\chi) \over 1+\chi} \ , \quad
   \tilde\chi = \chi + {\sqrt{1 - 4w} \over w(2+w)} \ .
  }}
 With the help of \wDef\ one can show that 
  \eqn{sExtra}{
   s(\tilde{e},\tilde{j};N) = s(e,j;N) 
    \left( {1+\chi \over 1+\tilde\chi} \right)^{5/2}
    \left( {\tilde\chi \over \chi} \right)^{1/2} \ .
  }

\begin{figure}[t]
   \vskip0cm
   \centerline{\psfig{figure=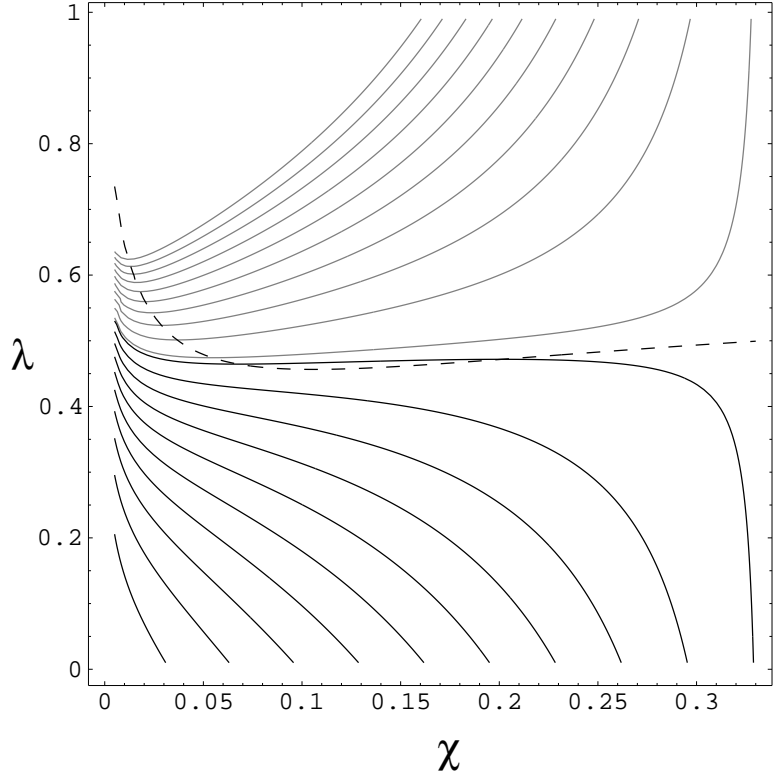,width=4in}}
   \vskip0.7cm
 \caption{The solid black lines are contours of constant $\chi_{av}$ with
$\chi_{av} < 1/3$.  The grey lines are contours of constant $\chi_{av}$
with $\chi_{av} > 1/3$.  The dashed line is the solution set of \eno{EntExt}.
The region below the dashed line is where entropy along a contour increases
as one moves to larger $\chi$.  If a contour intersects the dashed line,
then the maximum of the entropy function along that contour is at the first
intersection with the dashed line; otherwise the maximum is at the lower
border $\lambda = 0$.}\label{figA}
  \end{figure}

Because of the overall scale invariance of the theory, it is
equivalent to extremize
  \eqn{sigmaAv}{
   \sigma_{av} = {s_{av}^4 \over e_{av}^3}
  }
 with fixed
  \eqn{chiAv}{
   \chi_{av} = {27 \pi^2 \over 8 N^2} {j_{av}^4 \over e_{av}^3} \ .
  }
 Using \ejConstraint, \wDef, and \sExtra, one can obtain $\sigma_{av}$ and
$\chi_{av}$ explicitly as functions of $\chi$ and $\lambda$.  The domain of
$\chi$ is $[0,1/3]$ because it is by assumption the ratio pertaining to the
stable component, and the domain of $\lambda$ is $[0,1]$.  We have verified
numerically\footnote{Plots and numerics were obtained using Mathematica.}
that the contours of constant $\chi_{av}$ can be written as single-valued
functions, $\lambda = \lambda_{\chi_{av}}(\chi)$.  Thus the entropy along a
given contour is extremized when
  \eqn{EntExt}{
   \left( {\partial\sigma_{av} \over \partial \chi} \right)_{\chi_{av}} = 
   \left( {\partial\sigma_{av} \over \partial \chi} \right)_{\lambda} -
   \left( {\partial\sigma_{av} \over \partial \lambda} \right)_{\chi} 
   {\left( {\partial\chi_{av} / \partial \chi} \right)_{\lambda} \over
    \left( {\partial\chi_{av} / \partial \lambda} \right)_{\chi}} = 
    0 \ .
  }
 In figure~\ref{figA} we show contours of constant $\chi_{av}$ cut by
a dashed line indicating the solution set of \EntExt.  

\begin{figure}
   \vskip0cm
   \centerline{\psfig{figure=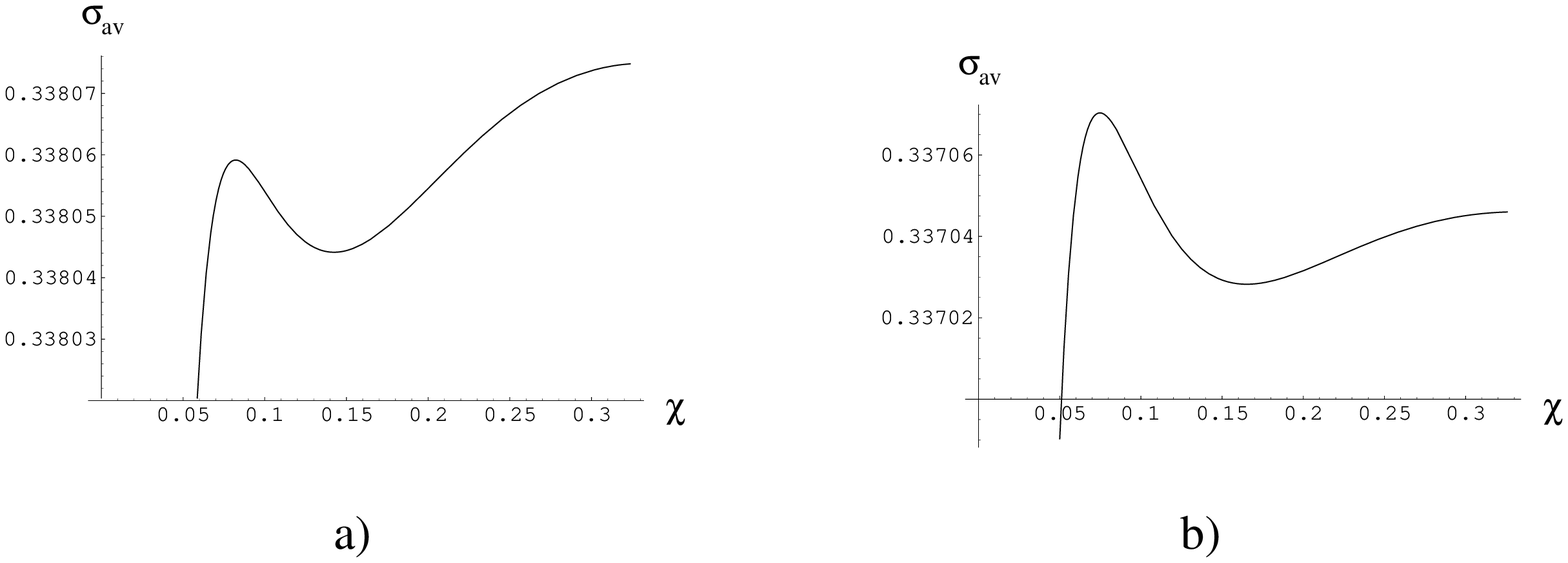,width=5in}}
   \vskip0.7cm
 \caption{The dimensionless ratio $\sigma_{av} = s_{av}^4/e_{av}^3$ is
plotted against $\chi$ for two values of $\chi_{av}$ on either side of the
phase transition: in a) $\chi_{av} = 0.324$ while in b) $\chi_{av} =
0.326$.  In a) we see that the $\chi = 0.324$, $\lambda = 1$ solid phase wins
out entropically; in b) the mixed phase with $\chi \approx 0.075$ and
$\lambda \approx 0.46$ is entropically favored.}\label{figB}
  \end{figure}

From figure~\ref{figA} one can conclude that for $\chi_{av} > 1/3$ the
entropically preferred state, among states with uniform temperature and
voltage, is a mix of the stable and unstable phases, with $\lambda \gsim
0.46$ and $\lambda \to 1$ as $\chi_{av} \to \infty$.  The $\chi_{av} \lsim
0.322$ mixtures are always entropically disfavored: for such $\chi_{av}$
the preferred state has $\lambda = 0$.  There is a narrow band of values,
$0.322 \lsim \chi_{av} < 1/3$, where there are two far-separated competing
states (see figure~\ref{figB}): one has $\lambda = 0$ and one has $\lambda
\approx 0.46$.  At $\chi_{av} \approx 0.325$ there is a ``transition''
where $\lambda$ jumps from $0$ to $0.46$.  The temperature and voltage also
jump at this transition.

We regard this as preliminary evidence that there is a first order phase
transition at a value of $\chi_{av}$ strictly less than $1/3$.  The
critical behavior associated with the boundary of stability would be cut
off finitely below the critical angular momentum by phase separation.  A
more definitive treatment would be possible if one could identify a
thermodynamically stable phase into which the system falls when the angular
momentum density is sufficiently high.  Our treatment has assumed that this
new stable phase is very slow to form compared to the time-scale on which
small inhomogeneities in $j$ can nucleate.  This is reasonable if the new
stable phase is some version of a multi-center solution, because passing to
such a phase would violate classical singularity theorems.

In \cite{cai} it was proposed that in the canonical ensemble, stability
persisted up to $\chi \approx 1.26$.  As a general rule we agree that
thermodynamic stability of a system depends on its environment.  However
the results of this section indicate that the instability of spinning
D3-branes at $\chi = 1/3$ is not easily avoided by changing ensembles.
Consider a state with total $e$ and $j$ arranged so that $\chi_{av}$ is
just slightly larger than $1/3$.  The pure phase state would correspond to
where one of the grey lines in figure~\ref{figA} intersects the top of the
frame.  At this point, $\tilde\chi = \chi_{av}$.  This is the sort of state
that the authors of \cite{cai} regard as stable in the canonical ensemble
(though not in the grand canonical ensemble).  However, the system may
increase its entropy continuously by flowing down along one of the grey
lines, until it reaches the dashed black line, where entropy is a maximum
at least among two-component mixed phase states of the prescribed
$\chi_{av}$.  In other words, the putatively stable state represents only a
saddle point of the entropy, not even a local maximum.  This is in contrast
with states with $0.325 \lsim \chi_{av} < 1/3$, where a system in a pure
phase state can increase its entropy only by making a big leap, all the way
along one of the fine black lines from the bottom of the frame to the
 \noindent
 leftmost intersection of the fine black line with the dashed black line.
In a nutshell, the pure phase states represented by the bottom of the frame
in figure~\ref{figA} are metastable against separation into a two-component
mixture, but the pure phase states represented by the top of the frame are
not even metastable.  These statements are independent of the environment
because we are talking only about shifting energy around within the
world-volume of the D3-branes.  The key point is that D3-branes have
spatial extent, so any little bit of it can be regarded as a ``system'' in
thermal contact with a ``heat bath'' consisting of the rest of the brane.
That is what makes the grand canonical ensemble relevant.  With black holes
the story is different: because there is no spatial extent, it is doubtful
that one can regard one part of the black hole as a heat bath for another
part.  If the D3-brane world-volume were compactified on a scale smaller
than the thickness of a domain wall between phases, then the situation
would be physically indistinguishable from a black hole, and calculations
in the canonical ensemble might become relevant.

\section{Euclidean Spinning Branes}
\label{EuclideanBranes}

\subsection{Spin and thermal boundary conditions}
\label{SpinBoundary}

Euclidean black brane solutions can be obtained from the Minkowskian ones
by sending $t \to -it$ and $\ell_j \to i \ell_j$, where the $\ell_j$ are
the angular momentum parameters that enter into the solutions of
\cite{CvYOne}.  Complexifying them is necessary in order to keep the $dt
d\phi$ parts of the metric real.  The consequences for the thermodynamics
can all be reasoned out from the fact that the angular momentum becomes
complex.  In brief, one can retain equations \MCSOne\ through \DiffRel\
with the replacements $y_j \to i y_j$ and $x_j \to i x_j$, but $y_H \to
y_H$ with no factor of $i$.  These replacements have a profound effect on
the physics: as we shall now argue, the phase space is $|x_j| < 1$, and
points on the boundary correspond to supersymmetric periodicity conditions
in the Euclidean path integral.

The new version of the first equation in \DiffRel\ is 
  \eqn{YHEDef}{
   y_{H,E}^{3-D} = \prod_j (1 - x_j^2) \ .
  }
 This relation becomes singular when any of the $x_i$ reaches $1$ in
absolute value, so the phase space is indeed the cube $|x_i| <
1$.  The new version of \OmegaTRatio\
and \pyFunct\ is
  \eqn{OTERatio}{
   {\Omega_j \over T} = 2\pi i {\partial y_H \over \partial y_j} = 
     -2\pi i {x_j \over 1-x_j^2} 
     {2 \over 3-D-\sum_k {2 x_k^2 \over 1-x_k^2}} \ .
  }
 We now want to study the limits of \OTERatio\ as we approach the boundary
of the cube.  Consider fixing some number of the $x_j$ at values finitely
far from the horizon and then letting all the others approach the horizon
at the same rate: for these others, say $x_j = s_j (1-\epsilon)$ where $s_j
= \pm 1$ and $\epsilon \to 0$.  For convenience define $s_j = 0$ for all
the $x_j$ whose values are fixed in the interior.  Let the number of
$x_j$'s which are approaching the boundary be $n$.  Approaching a generic
point on the boundary corresponds to $n=1$; $n>1$ means one is approaching
a point on an edge or a corner of the cube rather than a face.  It is
simple to see from \OTERatio\ that
  \eqn{OTEApproach}{
   {\Omega_j \over T} \to {2\pi i \over n} s_j \ .
  }
 For example, if we send $x_1 \to 1$ and hold all the other $x_i$ fixed,
then $\Omega_1/T \to 2\pi i$ and all other $\Omega_j/T \to 0$.

To understand what is happening in the world-volume theory, let us first
recall that in the absence of rotation, the usual thermal partition
function, $Z = \tr e^{-\beta H}$, can be computed by a Euclidean path
integral with period $1/T$ for Euclidean time.  Boundary conditions on
fermions are anti-periodic.  To add real angular momentum---the kind that
we have studied in all the previous sections---one would introduce a
chemical potential $\vec{\Omega}$ for particles with R-charge: acting on a
state $|\vec{\alpha}\rangle$ whose R-charge is specified by the vector
$\vec{\alpha}$ in the weight space of the R-symmetry group, the hamiltonian
$H$ would be replaced by $H - \vec{\alpha} \cdot \vec{\Omega}$
(cf.~\XiInt).  But in this section we are studying Euclidean angular
momentum, which corresponds to pure imaginary $\vec{\Omega}$.  Thus
  \eqn{RAct}{
    e^{\vec{\alpha} \cdot \vec{\Omega}/T} |\vec{\alpha}\rangle = 
      e^{\vec{H} \cdot \vec{\Omega}/T} |\vec{\alpha}\rangle = 
      R |\vec{\alpha}\rangle \ ,
  }
 where $\vec{H} = (H_1,H_2,\ldots,H_{\lfloor (D-1)/2 \rfloor})$ is a basis
of generators for the Cartan subalgebra.  $R = e^{\vec{H} \cdot
\vec{\Omega}/T}$ is an element of the compact R-symmetry group $SO(D-1)$
(or more precisely its covering group) when $\vec{\Omega}$ is pure imaginary.
The net effect of turning on $\vec{\Omega}$, as we see from \RAct, is to insert
$R$ into the partition function: instead of $Z = \tr e^{-\beta H}$, now we
have $Z = \tr R e^{-\beta H}$.  In the Euclidean path integral, $R$ just
specifies a twist in $SO(D-1)$ which one performs on all the fields before
identifying from $t_E = 1/T$ to $t_E = 0$.  This twist modifies the usual
(thermal) boundary conditions according to the R-charge of each field.

If we set $\Omega_1/T = 2\pi i$ and all other $\Omega_j/T = 0$, then $R =
(-1)^F$, where $F=0$ for bosons and $1$ for fermions.  That is, we recover
supersymmetric boundary conditions.  Maximal Poincar\'e supersymmetry is
recovered in this case because $R \eta = -\eta$ for any spinor $\eta$ of
$SO(D-1)$.  This {\it does not} mean that conformal invariance is
recovered.  In fact it is not: the near-horizon geometry is not anti-de
Sitter for Euclidean spinning branes, even in the large spin limit we are
considering.  We are led to conclude that the geometry must be describing a
physical state analogous to those considered in \cite{klt}, where Higgs
VEV's break conformal invariance.\footnote{Related ideas have recently been
explored in \cite{fgpwTwo}, although there are subtleties with the Wick
rotation in comparing with the present work.}  

For $n>1$, we can judge whether any part of supersymmetry is restored by
determining whether there are any spinors $\eta$ of $SO(D-1)$ which satisfy
$R\eta = -\eta$, where $R$ is the element of the covering group of
$SO(D-1)$ specified by $\Omega_j/T = 2\pi i/n$ for $1 \leq j \leq n$.  It
is straightforward to verify in this way that a fraction $(1/2)^{n-1}$ of
supersymmetry is preserved in all cases except $n=4$ for the M2-brane,
where $1/4$ of supersymmetry is preserved.

By studying Killing spinor equations in the spinning brane supergravity
geometries one can confirm the presence of unbroken supersymmetry.  This
was essentially done in \cite{DuffLiu} for the M2-brane.  To be more
precise, \cite{DuffLiu} includes an investigation of the Killing spinor
equations for BPS-saturated R-charged black holes of $N=8$ $D=4$ gauged
supergravity.  The results of \cite{Myers,CvGu1} indicate that Kaluza-Klein
reduction of spinning M2-branes leads precisely to large R-charged black
holes of the type studied by the authors of \cite{DuffLiu}.  They found
that $2^{5-n}$ supersymmetries are preserved by black holes with $n=1$,
$2$, or~$3$ charges, while adding a fourth charge does not break any
additional supersymmetry.  This counting is in agreement with the previous
paragraph: for example $n=1$ corresponds to $16$ supersymmetries, which is
the maximum when conformal symmetry is broken by the physical state.

The weight vectors listed in \WeightVectors\ can be used to determine the
allowed momenta $k_E$ around the Euclidean $S^1$.  These momenta determine
tree-level masses in the dimensionally reduced theory.  For $n<4$ there is
Bose-Fermi degeneracy in all cases (M2, D3, and M5). (See also figure~\ref{figD}.) The M2-brane with
$n=4$ exhibits no Bose-Fermi degeneracy.  This may be an indication that
our description of the world-volume theory of many M2-branes in terms
of excitations with the same $SO(8)$ quantum numbers as for a single
M2-brane is too simplistic to capture even the rough outlines of the field
theory.  

In order to further shed light on the nature of supersymmetry restoration  in
field theory and to make comparison with  the supergravity   results, let us turn
to the study of  the thermodynamic quantities.

We obtain $\Omega_j/T = (2\pi i)/n$ exactly when $x_i = \pm 1$ ($i=1,\cdots ,n$),
 but other thermodynamic relations become singular in this limit.
So we choose $\omega_i=(\pm
){2\pi i}/ [n (1-\epsilon)]$ ($x_i=\pm(1-\epsilon)$) ($i=1,...,n$,
$\epsilon\ll 1$), and for the sake of simplicity we set other $\omega_j$'s
to zero.  This is still not the most general way one could imagine
approaching the boundary of phase space: for instance, one could set $x_i =
\pm (1-r_i \epsilon)$ with arbitrary but fixed $r_i$.  This does not lead
to any more general scaling behavior than what we will observe, but the
coefficients in equations \eno{d3ft} through \eno{m2sgt} and (\ref{jerm})
 would change. 

First we compute the grand potential for the 
 field theory 
obtained from \XiInt\ with the supersymmetric particle content given in
\WeightVectors.  The field theory results in the $\epsilon \to 0$ limit are as follows:
 \begin{eqnarray}
{\rm D3:}& &\ \ \  {\xi\over T^4}= -\pi^2\left(\textstyle{2\over 3}\epsilon^3\ ,
    \textstyle{1\over 2}\epsilon^2\ , 
              \textstyle{4\over {27}}\epsilon\ \right), \  \  n=(1,2,3)\ ,
\label{d3ft}
\\
{\rm M5:}& &\ \ \  {\xi\over T^6}=-\pi^3\left(\textstyle{1\over 3}
\epsilon^4 \ ,
\textstyle{1\over {12}}\epsilon^2\, \right)\ ,  \ \ \ n=(1,2)\ , 
\label{m5ft}\\
{\rm M2:}& &\ \ \ {\xi\over T^3}=-\left(\pi\ln{2}\  \epsilon^2\ ,  
              \pi(\textstyle{3\over 2}+\ln{2\over{\pi\epsilon}})
	      \epsilon^2\ ,
              2.0988
	      \epsilon\ , 
   0.33480\,\right)\ ,  \ \ \ n=(1,2,3,4)\ .
 \label{m2ft}\end{eqnarray}
For a single
M5-brane, or D3-brane  this field theory analysis is  sound: it amounts simply
to counting states in a free field theory.  (It is not even plagued by the
divergences that we used the principle value prescription to solve in
section~\ref{FTanal}: those divergences were for $\omega_i$ real, and are
completely avoided by pure imaginary $\omega_i$.) 
 For $N>1$ our understanding of the field theory at finite temperature is
limited.  The naive approach is to hope that it has some properties in
common with $O(N^3)$ copies [$O(N^2)$ copies] of the $N=1$ theory in the
M5- [D3-] brane.  This approach is not well motivated physically, but it
does give the correct scalings with $N$ and $\epsilon$ as predicted by
supergravity.
 The field theory on multiple M2-branes not very well understood.  It is
based on a particular infrared limit of the D2-brane gauge theory where
$SO(8)$ symmetry is thought to be recovered (see for example
\cite{SethiSusskind,SeibergSixteen}).  For a single brane, a simple Hodge
dualization of a vector field into a scalar field suffices to show this,
but for multiple branes there does not seem to be a simple argument
directly from field theory.

The positive powers of $\epsilon$ in \eno{d3ft}-\eno{m5ft} appear because
of the Bose-Fermi degeneracy in the spectrum of allowed $k_E$.  This is
substantially the only indicator of supersymmetry in this free field theory
computation.

Let us now compare the field theory results to those obtained in
supergravity. There the grand potential $\xi$ is given by \GrandPot.  In
the $\epsilon \to 0$ limit, $\xi$ has the following scaling behavior:
  \begin{equation}
  \xi\sim T^{p+1} \epsilon^{{{n(p+1)}\over{3-D}}+p+1}\ . 
 \label{gpe} \end{equation}
 The specific results  are:
\
\begin{eqnarray}
{\rm D3:}& &\ \ \ {\xi\over T^4}=-\pi^2\left(16\epsilon^{3} \ ,
            \textstyle{1\over 2}\epsilon^2\ ,
	    \textstyle{4\over{81}}\epsilon\ \right)\ , \ n=(1,2,3) \ ,
	    \label{d3sg}\\
{\rm M5:}& &\ \ \ {\xi\over T^6}=-\pi^3\left(  \textstyle{16\over 3}\epsilon^{4}
\ ,  \textstyle{1\over {48}}\epsilon^2\ \right)\ ,\ \ \ 
	       n=(1,2)\ ,
\label{m5sg}
\\
{\rm M2:}& &\ \ \ {\xi\over T^3}= -\pi^2\left(\textstyle{64\over{3}}\epsilon^{5/2}\ ,
               \textstyle{{4\sqrt{2}}\over {3}}\epsilon^{ 2}\ ,
 \textstyle{32\over 81}\epsilon^{3/2}\ ,\textstyle{1\over
{6\sqrt{2}}}\epsilon
 \ \right)\ , \  n=(1,2,3,4) \ . 
 \label{m2sg} \end{eqnarray}

It is instructive to trace the behavior the inverse temperature in this limit.
Using the expression  \Beta\ one obtains:
\begin{eqnarray}
{\rm D3:}& &\ \ \ \beta\equiv {1\over T}={{2\pi}\over n} {R^2\over
r_0}(2\epsilon)^{1-{n\over  4}}\ , \ n=(1,2,3) \ ,
	    \label{d3sgt}\\
{\rm M5:}& &\ \ \ \beta\equiv {1\over T}={{2\pi}\over n} \left({R^{3}\over
r_0}\right)^{1\over 2}(2\epsilon)^{1-{n\over 3}}\ ,\ \ \ 
	       n=(1,2)\ ,
\label{m5sgt}
\\
{\rm M2:}& &\ \ \ \beta\equiv {1\over T}={{2\pi}\over n} {R^3\over
r_0^2}(2\epsilon)^{1-{n\over 6}}\ , \  n=(1,2,3,4) \ . 
 \label{m2sgt} \end{eqnarray}

Note that for finite temperature to remain finite  $r_0\sim (\epsilon^{1-{n\over 4}}\ ,
\epsilon^{{6-2n}\over 3}\ ,\epsilon^{{6-n}\over {12}}) $ for D3-, M5- and
M2-brane, respectively.

Comparing  the supergravity results  \eno{d3sg} through \eno{m2sg}
 with  those the field theory  analysis (\eno{d3ft} through \eno{m2ft})
  reveals that
   except for rational prefactors the supergravity and field theory
results are in good  qualitative agreement for both D3- and M5-brane, as
expected since a single D3- and M5-brane field theory analysis is believed to be
sound.
On the other
hand there is a  qualitative discrepancy between the two approaches for 
the M2-brane case, again indicating that the field theory of M2-brane is poorly
understood.

Most importantly, we see that the  energy density $e=-p\xi$ scales in
{\it all cases} with a positive power of $\epsilon$.  Recall that $e$ is
the energy density over and above the energy density of a BPS-saturated
brane configuration.  By the same argument that zero energy states in a
supersymmetric theory are supersymmetric states, we conclude that a state
with $e=0$ preserves at least some of the original supersymmetries of the
brane.

We would like to conclude this subsection by commenting on the supergravity
analysis of the spectrum of the theory.  This will become relevant in our
discussion in the next subsection of models of confinement based on
spinning branes.  Let us concentrate on a minimally coupled scalar $\phi$,
that is a scalar whose linearized equation of motion is the Laplace
equation in the background of the Euclidean spinning brane:
 \begin{equation}
{1\over{\sqrt{-g}}}\partial_\mu(\sqrt{-g}g^{\mu\nu})\partial_\nu \phi=0.
\end{equation}
With the    ansatz $\phi=e^{ikx}\chi(r)$,  where  
 $x$ is a world-volume direction,  and (for the
sake of simplicity) $\chi(r)$ depends only on the radial coordinate $r$,
 one obtains a radial 
 equation of the following form:\footnote{In deriving 
 the above equation the full metric of the spinning brane has to be
used. We used the  metric of the general $D$-dimensional 
rotating two-charge black hole solutions  given in  \cite{CvYOne}.
Again, the corresponding spinning  (D3, M5, M2)-branes are 
obtained by turning off one of  the two  charges of  the ($D=7,6,9$-dimensional) black
hole, and 
lifting the solution  to ten and eleven dimensions,
respectively.}
\begin{eqnarray}
D\ \ {\rm odd}:& &\ \ \ {1\over r
R^{D-3}}\partial_r\left[ r^{-1}\prod_{i=1}^{\left\lfloor {{D-1}\over
2} \right\rfloor}(r^2-\ell_i^2)-2mr \right]\partial_r \chi(r)=k^2\chi(r)\ , \nonumber\\
D\ \ {\rm even}: & &\ \ \ {1\over r
R^{D-3}}\partial_r\left[ \prod_{i=1}^{\left\lfloor {{D-1}\over
2} \right\rfloor}(r^2-\ell_i^2)-2mr \right]\partial_r \chi(r)=k^2\chi(r)\ . 
\end{eqnarray}
These wave equations were derived in \cite{rs} 
and \cite{crst} for the D3-brane and the M5-brane, respectively.  
In the field theory, the values of 
$|k|$ will be masses of $O^{++}$ glueballs.

In terms of the notation introduced in this paper these wave equations assume the
form:
 \begin{eqnarray}
D\ \ {\rm odd}:& &\ \ \ { x}^{-1}
\partial_x\left[ y_H^{D-3}x^{-1}\prod_{i=1}^{\left\lfloor {{D-1}\over
2} \right\rfloor}(x^2-x_i^2)-x \right]\partial_x \chi=y_H^2(kr_0)^2\left({R\over
r_0}\right)^{D-3}\chi\  ,\nonumber \\
D\ \ {\rm even}: & &\ \ \ x^{-1}
\partial_x\left[ y_H^{D-3}\prod_{i=1}^{\left\lfloor {{D-1}\over
2} \right\rfloor}(x^2-x_i^2)-x \right]\partial_x \chi=y_H^2(kr_0)^2\left({R\over
r_0}\right)^{D-3}\chi\ . 
\end{eqnarray}
 Our radial variable is $x = {r \over r_0 y_H}$.  The quantities $x_i$,
$y_i$, $y_H$, $r_0$, and $R$ were introduced at the beginning of
section~\ref{SUGRAThermo}, but because we are now operating in Euclidean
space we must recall the replacements $x_j \to i x_j$ and $y_j \to i y_j$.
When $n$ equal Euclidean angular momenta becoming large at fixed energy
(that is, $x_i \to \pm (1-\epsilon)$ for $i=1\ldots,n$), the limiting
behavior for $k$ is 
\begin{equation}
k\equiv M_{\rm glueball}={\textstyle{{2\pi}\over n}} {\cal Q} (2\epsilon)^{1-{n\over 2}}T\ , \ \
(n=1,\cdots , {\textstyle{{[D-1]}\over 2}})\ , 
\label{jerm}\end{equation} 
where $T$ is the Hawking temperature  \Beta\  (which at the boundary takes the
values (\ref{d3sgt})-(\ref{m2sgt})), and ${\cal Q}$ is a pure number
determined by the following eigenvalue equation:
 \begin{eqnarray}
D\ \ {\rm odd}:& &\ \ \
{x}^{-1}\partial_x\left[ x^{-1}\prod_{i=1}^{\left\lfloor {{D-1}\over
2} \right\rfloor}(x^2-x_i^2) \right]\partial_x \chi=-{\cal Q}^2\chi
  , \nonumber\\
D\ \ {\rm even}: & &\ \ \ x^{-1}
\partial_x \left[ \prod_{i=1}^{\left\lfloor {{D-1}\over
2} \right\rfloor}(x^2-x_i^2) \right] \partial_x \chi=-{\cal Q}^2\chi\ , 
\label{CalN}\end{eqnarray}
where now $x_i^2=1$ $(i=1,\ldots , n)$ and other $x_j$'s are set to
zero. (As before, 
with  non-zero $x_j^2\ne 1$'s and $x_i=\pm (1-r_i\epsilon)$, the 
  equations \eno{CalN} are of the same form, and  the mass gap  
  \eno{jerm} has the same scaling behavior with $\epsilon$, but the actual numerical
  prefactors in \eno{jerm} change.)  Discrete eigenvalues ${\cal Q}$ arise from 
demanding that $\chi$ is normalizable.\footnote{Certain
cases of \eno{CalN} can be solved in terms of hypergeometric
functions---for instance $n=1$, $D=7$---and exact expressions for the
eigenvalues ${\cal Q}$ were obtained in \cite{fgpwTwo}.  Some other cases
can be solved in terms of Heun functions.}

Thus $M_{gap}\sim \sqrt{\epsilon}T$ for $n=1$, $M_{gap}\sim T$ for $n=2$,
and $M_{gap}$ diverges for $n > 2$.  For $n>2$ it is possible that there is
actually no discrete spectrum in the $\epsilon \to 0$ limit.  This point is
under investigation.

The mass gap  can also be quantified
in terms of the ratios  of the  angular momentum densities at
the boundary,  $j_i$, ($i=1,\cdots , n$) and  $e$.
 It is straightforward to obtain the following
scaling behavior:
$j_i\sim T^{p}\epsilon^{{{n(p+1)}\over{3-D}}+p}$.  One can easily show that 
\begin{equation}
M_{gap}\sim \left({j_i^2\over e}\right)^{1/(p-1)}\sim T 
\epsilon^{1-{n(p+1) \over (D-3)(p-1)}}=T\epsilon^{1-{n\over 2}}\ . 
\label{jer}\end{equation}

Some of the observations we have made about the scaling of the mass gap and
about supersymmetry restoration also follow easily from the results of
\cite{russo,cort,rs,crst}.  The connection that does not seem to have been
made there is that the field theory itself recovers supersymmetry in the
limit of small $\epsilon$ because of the insertion of an element of the
R-symmetry group into the partition function.

\subsection{Models of confinement}
\label{Confine}   

The usual route to $\hbox{QCD}_4$ \cite{witHolTwo} is to compactify the
M5-brane worldvolume on a circle twice.  The first time, one uses
supersymmetric boundary conditions on a circle of circumference $2\pi R_1$,
and by the M-IIA duality the result is a D4-brane which, like the M5-brane,
is near-extreme.  The second compactification is applied to the Euclidean
time coordinate of the D4-brane: it has period $1/T$, and we impose thermal
rather than supersymmetric boundary conditions.  This second
compactification is the important step.  The five-dimensional theory one is
starting with on the D4-brane is the Euclidean version of the $SU(N)$ gauge
theory with maximal supersymmetry which descends from the ten-dimensional
${\cal N}=1$ supersymmetric Yang-Mills theory.  The resulting theory in
four dimensions will also have $SU(N)$ gauge fields, but no supersymmetry,
and it is the candidate for $\hbox{QCD}_4$.  Tracing through the M-IIA
relationship determining the string coupling in terms of the
compactification radius $R_1$, and then the relationships between the string
coupling, the D4-brane gauge coupling, and the four-dimensional
$\hbox{QCD}_4$ coupling $g_{YM}^2$, one finds $g_{YM}^2 = 2\pi R_1 T$.  The
relevant points for our analysis are: 1) the compactification process is
unchanged when there is angular momentum in the directions orthogonal to   
the M5-brane; 2) the 't~Hooft coupling $\lambda = g_{YM}^2 N$ needs to be  
large for the supergravity to be valid.

When there is no angular momentum, the fermions are anti-periodic in the
Euclidean time used in the $\hbox{D4} \to \hbox{QCD}_4$ compactification,
so in the four-dimensional lagrangian they have masses $\pi T$.  It is
usually assumed that at large 't~Hooft coupling, the scalars pick up
comparable masses.  From supergravity one learns that the mass gap is also
of order $T$.  This is sensible from a field theory point of view: in the
renormalization group flow of a gauge theory in which the 't~Hooft coupling
is large at energies comparable to the masses of the matter fields, one
expects confinement at only lower energies.  This is to be contrasted with
a gauge theory where the 't~Hooft coupling is weak at energies comparable
to the masses of the lightest matter fields.  There we have the standard
one-loop relation $M_{\rm gap} \sim \exp\left( {-{8\pi^2 \over b_0
g_{YM}^2}} \right) m_{\rm matter}$, and $b_0 = {11 \over 3} N$.\footnote{We
thank O.~Aharony for pointing out to us the comparison with the one-loop
analysis.}

\begin{figure}[t]
   \vskip0.5cm
  \centerline{\psfig{figure=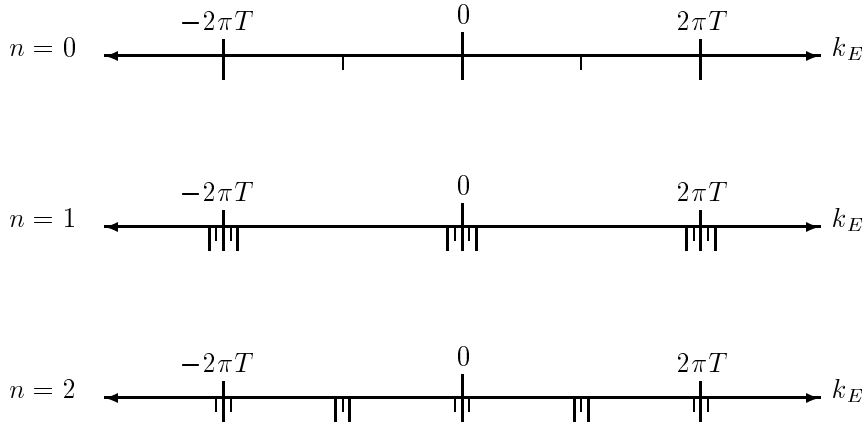,width=4.5in}}
   \vskip0.5cm
 \caption{The spectrum of momenta in the compactified $S^1$ direction of
the D4-brane world volume.  The larger ticks indicate bosons; the smaller
ticks indicate fermions.  The first line follows from standard thermal
boundary conditions.  The second and third lines show what happens when one
approaches the boundary of the phase space (large angular momentum for
fixed energy), either at an edge ($n=1$) or a corner ($n=2$).  The
splittings in these cases are on the order $\epsilon T$.  In the close
groupings of ticks in the second and third lines, the degeneracies follow
the ratios $1:4:6:4:1$ and $1:2:1$, respectively.}\label{figD}
  \end{figure}

When there is a large angular momentum, or, in the parlance of
section~\ref{SpinBoundary}, small $\epsilon$, the story is rather different.
The fermion masses are not all on the order of the temperature: as
indicated in figure~\ref{figD}, some of them are on the order of $\epsilon
T$.
 Supersymmetry is restored in the limit $\epsilon \to 0$.  In the $n=1$
case, for energies much smaller than $T$, only the center grouping of modes
is available, and the theory is four-dimensional ${\cal N}=4$
super-Yang-Mills.  In the $n=2$ case, the center group of states fills out
the ${\cal N}=2$ super-Yang-Mills multiplet.

Supergravity calculations which we summarized in the previous subsection
indicate that the scaling of the mass gap is $M_{\rm gap} \sim
\sqrt{\epsilon} T$ for $n=1$ and $M_{\rm gap} \sim T$ for $n=2$.  Near the
end of this section we explain why this scaling is rather at odds with our
field theory expectations.  For now we will regard the scaling of $M_{\rm
gap}$ as a non-trivial prediction of supergravity.

In view of the existence of matter fields comparable to or much lighter
than the scale of confinement, it is no surprise that large angular momenta
did not suffice to decouple the ``glueballs'' charged under the global
symmetry group which rotates these quarks among themselves \cite{cort}.
The existence of such ``Kaluza-Klein cousins'' has been a persistent
feature of all supergravity models of $\hbox{QCD}_4$, but it need not be
regarded as an artifact of the construction which must disappear in a full
string theory treatment.  In any confining gauge theory (including
real-world QCD) in which there are fundamental matter fields with some
flavor symmetry and with masses comparable to or lighter than the
confinement scale, one expects to have flavored hadrons as well as
flavor-neutral glueballs.  It may not be reasonable to hope that a full
string theory treatment of these backgrounds will decouple all matter
fields and leave us with pure glue.  It seems more likely that it will
leave us with confining glue coupled to light matter.

It makes sense that glueball mass ratios should be roughly the same
\cite{crst} when one makes one angular momentum large as when one makes two
large in any ratio except unity.  In terms of the $x_i$ variables discussed
in the section~\ref{SpinBoundary} and defined in \XDef, an M5-brane with
one angular momentum large with the other zero corresponds to $x_1 \to 1$
while $x_2 = 0$, whereas two large angular momenta with fixed ratio
corresponds to $x_1 \to 1$ with $x_2/x_1 = \eta$.  As long as $\eta \neq
1$, one is approaching the edge rather than the corner of the square which
represents the phase space of the spinning brane, and the pattern of
tree-level masses shown in figure~\ref{figD} obtains.  Glueball mass ratios
should be and are approximately independent of $\eta$ because the matter
fields whose tree-level masses are small in comparison with the confinement
scale.  When $\eta = 1$, the pattern of tree-level masses is as shown in
the $n=2$ line of figure~\ref{figD}.  We expect that glueball mass ratios
will change at this point because the relevant matter fields have changed
their tree-level masses.\footnote{The term ``glueball'' is somewhat of a
misnomer since we are thinking of states which should involve adjoint
matter fields even if they are singlets under the global flavor symmetry
that the matter fields carry.  The term ``glueball'' is in use because of
the expectation (unlikely in our view) that the supergravity models
discussed will correspond to pure $\hbox{QCD}_4$.}

One difficulty with the original proposal of \cite{witHolTwo} is that the
scale of the fifth dimension is the same as the scale of confinement.  Thus
any process whose energy is high enough to probe the gauge dynamics
non-trivially---in the sense that colored states, not just an effective
theory like soft pions, would be needed to analyze the process---would also
probe the fifth dimension.  One might therefore ask in what sense one has
defined a four-dimensional gauge theory at any scale.  The spinning branes
construction is better in this regard because for $n=1$ there is a large
range of energies, $\sqrt\epsilon T \ll E \ll T$, where the theory is
four-dimensional and unconfined.  The decoupling \cite{cort} of
``glueballs'' with Kaluza-Klein momentum in the fifth dimension is the
tangible evidence that the length of the fifth dimension is much smaller
than the inverse QCD scale.

Other approaches to confinement avoid a fifth dimension altogether by using
only D3-branes, either in type~0 theory \cite{minConfine} or with a varying
dilaton in type~IIB theory \cite{gDil} (see also \cite{KehagSfet}.  These
models have their own difficulties: in the former the gravity description
breaks down in the ultraviolet, and there is a tachyon whose mass is on the
string scale; while in the latter one seems to need some excited string
state to start the dilaton flowing, and asymptotic freedom cannot be
achieved.  Both approaches involve naked singularities in the bulk.  

\begin{figure}[t]
   \vskip0.5cm
  \centerline{\psfig{figure=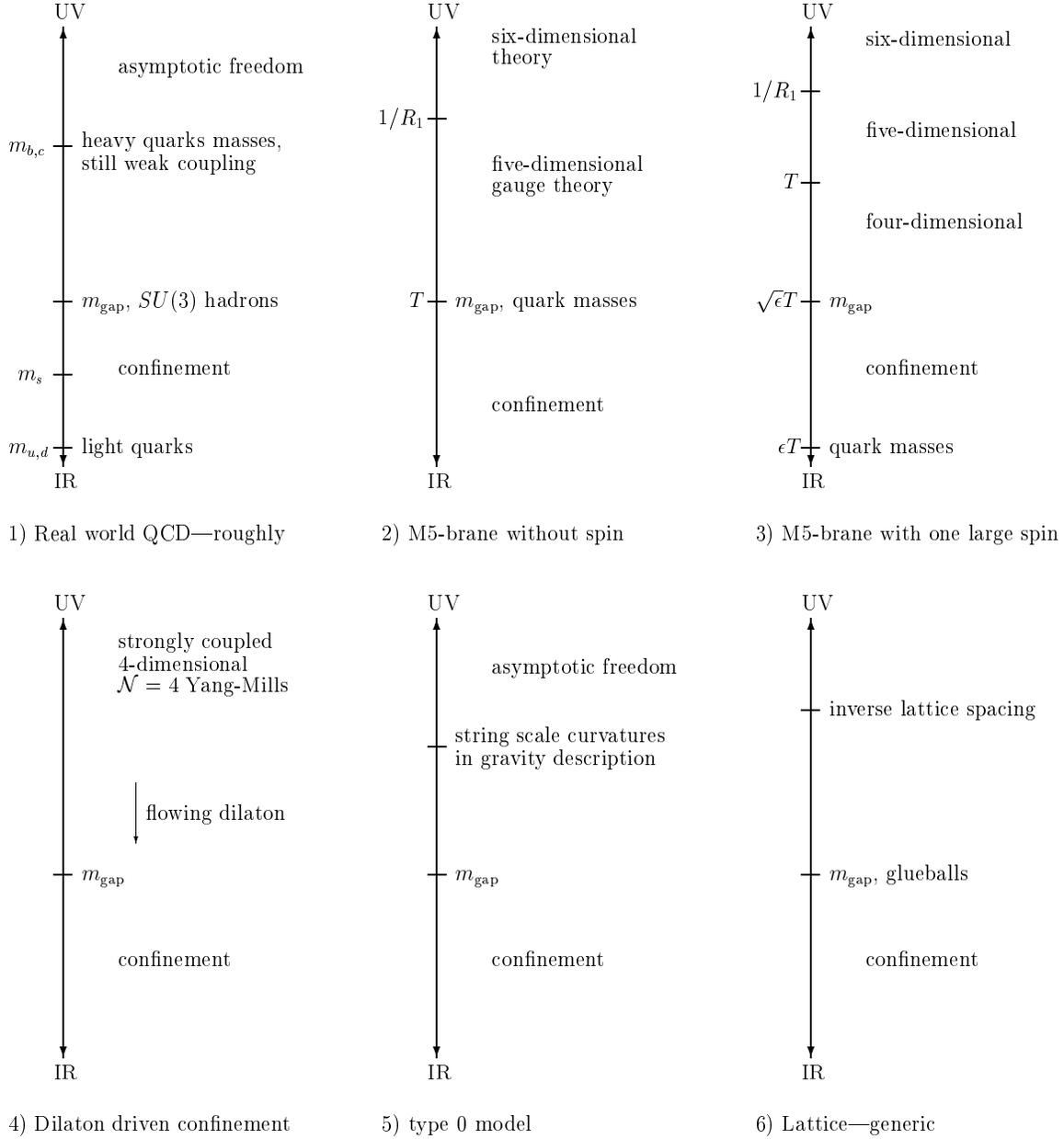,width=6in}}
   \vskip0.5cm
 \caption{Energies scales in various attempts to find a supergravity dual
to confining gauge theories.  Also included for purposes of comparison is a
cartoon sketch of real-world QCD and a generic lattice model.  The quark
masses shown are the parameters that enter into the lagrangian: constituent
masses.}\label{figC}
  \end{figure}

We summarize the taxonomy of holographic approaches to confining gauge
theories in figure~\ref{figC}.  The basic features of confinement are not
expected to depend on the existence of matter fields or on their masses,
provided that the beta function remains negative.  All the gravity models
have a string tension which is much larger than the square of the mass gap.
This leads to some behaviors very unlike real QCD \cite{gog}, in particular
nearly flat Regge trajectories.  In figure~\ref{figC}, we have indicated
only the mass gap---that is, the mass scale seen by closed string probes.
In all cases except 3), the QCD string tension would be located between
this mass scale and the next highest scale; in 3) the relation of the QCD
string tension and the temperature can only be determined given the
relative sizes of $\epsilon$ and the 't~Hooft coupling $g_{YM}^2 N$.

Since only one large angular momentum is needed to make the scalings we
have discussed, it should be possible to construct variants of the spinning
M5-brane approach by putting the branes at some spacetime singularity.
Orbifold backgrounds would be expected to lead to bifundamental quarks
rather than adjoints.  There might even be chiral models analogous to those
discussed in \cite{Uranga}.  A model with bona-fide light fundamental
quarks could be difficult to construct, since the fundamental
representation is the charge carried by one end of a string ending on a
D-brane, and not typically by an open string with both its ends on the
brane.

There is one outstanding difficulty: why for $n=1$ is
the confinement scale much higher than the tree-level masses of the light
matter fields ($\sqrt{\epsilon} T \gg \epsilon T$)?  The most obvious
resolution would be for at least some of the light masses to receive
radiative corrections on the order $\sqrt{\epsilon} T$ from diagrams where
heavy fields run around loops.
 \noindent
 If this were to happen, then superconformal
invariance would be broken at a scale on the order of $\sqrt{\epsilon} T$,
and confinement would follow at an energy scale only slightly lower.  But
the following simple estimate suggests that perturbative effects cannot
generate a sufficiently large radiative correction.  Scalar mass
renormalizations at one loop typically arise from integrals of the form
  \eqn{DimRegInt}{
   F(M) = \int {d^4 q \over (2\pi)^4} {1 \over q^2 + M^2}
        = {M^2 \over 8\pi^2} \log {M \over \mu} \ ,
  }
 where we have used dimensional regularization and the minimal subtraction
scheme.  $M$ is the mass of the heavy fields running around the loop.
Typically, in a theory with supersymmetry broken by mass splittings for the
heavy fields on the order $\Delta M$, we have $\delta m^2 \sim (\Delta M)
F'(M) \sim (\Delta M) M \log {M \over \mu}$ on account of the
near-cancellation of bosonic and fermionic contributions.  With mass
breakings such as the ones indicated in figure~\ref{figD}, the cancellation
is even better: $\delta m^2 \sim (\Delta M)^{\cal N} F^{({\cal N})}(M)$
where ${\cal N}$ is the degree of supersymmetry restored when $\Delta M =
0$.  In particular, $\delta m^2 \sim \epsilon^4 T$ for $n=1$, where ${\cal
N}=4$ supersymmetry is restored in the $\epsilon \to 0$ limit.

Although perturbation theory seems to have failed, we find it plausible
that the heavy fields should play a role in setting the scale of
supersymmetry breaking at $\sqrt{\epsilon} T$.  This scale is just the
geometric mean of the tree-level masses of the light and heavy fields, but
its relevance to the dynamics may emerge in field theory only through
features peculiar to strong coupling gauge interactions.

To summarize our discussion: In no sense should spinning M5-brane solutions
with a large angular momentum be regarded as the analog of improved lattice
actions for pure $\hbox{QCD}_4$.  The fermions' tree level masses are not
all far above the confinement scale.  In fact, some of them are far smaller
than this scale.  We do not understand the field theory mechanism that sets
the scale of supersymmetry breaking and confinement at $O(\sqrt{\epsilon}
T)$ for the case of one large angular momentum.  However, if we take this
scaling as a given, we do understand why glueball mass ratios should be
universal for two large unequal angular momenta and somewhat different when
they are equal: it is because the pattern of supersymmetry breaking is
universal in the first case, but different in the second.  It would be
delightful if the near-restoration of supersymmetry could be exploited to
make more definite predictions about the confining dynamics than is
possible for pure $\hbox{QCD}_4$.

The D3-brane with periodic Euclidean time has been proposed as a model for
pure $\hbox{QCD}_3$ \cite{witHolTwo}.  Without angular momentum, the bare
three-dimensional fermion masses and also the mass gap are on the order of
$T$.  With $n$ equal large Euclidean angular momentum, the light fermion
masses are $m_{\rm bare} \sim \epsilon T$, whereas the mass gap is $M_{\rm
gap} \sim \epsilon^{1 - {n \over 2}} T$ for $n=1$ or $2$.  
Again it is not easy to see what
in field theory sets this confinement scale.  
Universality of
glueball mass ratios can be understood in a similar way to the
four-dimensional case, but we regard it as a more pressing problem to
understand the overall scaling of the mass gap from a field theory
perspective. 

The supergravity analysis of the previous subsection applies to Euclidean
spinning M2-branes as well.  However, in view of the poor understanding of
the large $N$ field theory we refrain from drawing any further comparisons.

\section{Conclusion}
\label{conclusions}
 The purpose of this paper was to provide a comprehensive study of the
thermodynamics of spinning D3-, M5-, and M2-branes.  Our results address
the following aspects:
 \begin{itemize}
 \item {\it Thermodynamic Stability and Critical Behavior}
 
We derived general stability constraints for near-extreme D3-, M5- and
M2-branes, all with the maximum number of independent spins.  The region
where the entropy is subadditive as a function of both the energy and the
angular momenta is strictly smaller than the region where the specific heat
is positive.  To illustrate one of the possible instabilities that develops
when the entropy is not subadditive, we studied states where the angular
momentum is unevenly distributed over the world volume, and found them to
be entropically preferred over the uniform distribution despite the fact
that they are not themselves stable configurations.

\item {\it Field Theory}

We generalized the field theory model of \cite{gspin} to the case of
multiple R-charges; the dynamics is proposed to be described effectively by
a collection of weakly interacting massless ``quasi-particles'' whose
R-charges and spins are those of the abelian theory on a single brane.  If
one assumes in a addition that there is a nonzero ``permeability'' for
applied ``voltages'' dual to the various R-charges, then this model
reproduces the critical behavior of the supergravity analysis.

\item {\it Euclidean spinning branes}

Euclidean branes with at least one angular momentum density which is large
compared to the energy density restore some fraction of supersymmetry.  We
showed how this supersymmetry restoration arises in the field theory: an
element of the R-symmetry group is inserted in the partition function,
which for one large spin approaches $(-1)^F$.  The spectrum is nearly
supersymmetric, with the Bose-Fermi degeneracy broken by $\Delta
M\sim\epsilon T$, where $\epsilon$ is a dimensionless parameter that goes
to zero in the limit of large angular momentum.  The supergravity analysis
reveals that if there is confinement, it occurs on a scale
$M_{gap}\sim T\epsilon^{1-n/2}$
when $n$ equal Euclidean angular momenta are made large.  We have explained
how this picture elucidates some of the results of previous papers
\cite{cort,crst}.  We have left open the problem of identifying the field
theory mechanism that sets the scale of supersymmetry breaking.

\end{itemize}

\section*{Acknowledgements}

We thank D.~Gross, F.~Larsen, H.~L\" u, P.~Kraus, E.~Martinec,
A.~Strominger, and the participants of the Montreal Black Holes II Workshop
for useful discussions.  We also thank the JHEP referee for a thorough
review and useful suggestions.  The work of M.C.\ was supported in part by
U.S. Department of Energy Grant No.  DOE-EY-76-02-3071. The work of S.S.G.\
was supported by the Harvard Society of Fellows, and also in part by the
National Science Foundation under grant number PHY-98-02709.

\newpage

\bibliography{multi}

\begingroup\raggedright\begin{thebibliography}{10}

\bibitem{gspin}
S.~S. Gubser, ``Thermodynamics of spinning D3-branes,''
  \href{http://xxx.lanl.gov/abs/hep-th/9810225}{{\tt hep-th/9810225}}.

\bibitem{gl}
R.~Gregory and R.~Laflamme, ``Black strings and p-branes are unstable,'' {\em
  Phys. Rev. Lett.} {\bf 70} (1993) 2837,
  \href{http://xxx.lanl.gov/abs/hep-th/9301052}{{\tt hep-th/9301052}}.

\bibitem{juanAdS}
J.~Maldacena, ``The Large N limit of superconformal field theories and
  supergravity,'' {\em Adv. Theor. Math. Phys.} {\bf 2} (1998) 231,
  \href{http://xxx.lanl.gov/abs/hep-th/9711200}{{\tt hep-th/9711200}}.

\bibitem{gkPol}
S.~S. Gubser, I.~R. Klebanov, and A.~M. Polyakov, ``Gauge theory correlators
  from noncritical string theory,'' {\em Phys. Lett.} {\bf B428} (1998) 105,
  \href{http://xxx.lanl.gov/abs/hep-th/9802109}{{\tt hep-th/9802109}}.

\bibitem{witHolOne}
E.~Witten, ``Anti-de Sitter space and holography,'' {\em Adv. Theor. Math.
  Phys.} {\bf 2} (1998) 253, \href{http://xxx.lanl.gov/abs/hep-th/9802150}{{\tt
  hep-th/9802150}}.

\bibitem{cort}
C.~Csaki, Y.~Oz, J.~Russo, and J.~Terning, ``Large N QCD from rotating
  branes,'' \href{http://xxx.lanl.gov/abs/hep-th/9810186}{{\tt
  hep-th/9810186}}.

\bibitem{CvYInt}
M.~Cveti\v{c} and D.~Youm, ``Rotating intersecting M-branes,'' {\em Nucl.
  Phys.} {\bf B499} (1997) 253,
  \href{http://xxx.lanl.gov/abs/hep-th/9612229}{{\tt hep-th/9612229}}.

\bibitem{CvYOne}
M.~Cveti\v{c} and D.~Youm, ``Near BPS saturated rotating electrically charged
  black holes as string states,'' {\em Nucl. Phys.} {\bf B477} (1996) 449--464,
  \href{http://xxx.lanl.gov/abs/hep-th/9605051}{{\tt hep-th/9605051}}.

\bibitem{klt}
P.~Kraus, F.~Larsen, and S.~P. Trivedi, ``The Coulomb branch of gauge theory
  from rotating branes,'' \href{http://xxx.lanl.gov/abs/hep-th/9811120}{{\tt
  hep-th/9811120}}.

\bibitem{rs}
J.~Russo and K.~Sfetsos, ``Rotating D3 branes and QCD in three dimensions,''
  \href{http://xxx.lanl.gov/abs/hep-th/9901056}{{\tt hep-th/9901056}}.

\bibitem{crst}
C.~Csaki, J.~Russo, K.~Sfetsos, and J.~Terning, ``Supergravity models for
  (3+1)-dimensional QCD,'' \href{http://xxx.lanl.gov/abs/hep-th/9902067}{{\tt
  hep-th/9902067}}.

\bibitem{ktOne}
I.~R. Klebanov and A.~A. Tseytlin, ``Entropy of near extremal black p-branes,''
  {\em Nucl. Phys.} {\bf B475} (1996) 164--178,
  \href{http://xxx.lanl.gov/abs/hep-th/9604089}{{\tt hep-th/9604089}}.

\bibitem{CvGu1}
M.~Cveti\v{c} and S.~S. Gubser, ``Phases of R charged black holes, spinning
  branes and strongly coupled gauge theories,''
  \href{http://xxx.lanl.gov/abs/hep-th/9902195}{{\tt hep-th/9902195}}.

\bibitem{cai}
R.-G. Cai and K.-S. Soh, ``Critical behavior in the rotating D-branes,''
  \href{http://xxx.lanl.gov/abs/hep-th/9812121}{{\tt hep-th/9812121}}.

\bibitem{MerminWagner}
N.~D. Mermin and H.~Wagner, ``Absence of ferromagnetism or antiferromagnetism
  in one- dimensional or two-dimensional isotropic Heisenberg models,'' {\em
  Phys. Rev. Lett.} {\bf 17} (1966) 1133--1136.

\bibitem{gkSchwing}
S.~S. Gubser and I.~R. Klebanov, ``Absorption by branes and Schwinger terms in
  the world volume theory,'' {\em Phys. Lett.} {\bf B413} (1997) 41--48,
  \href{http://xxx.lanl.gov/abs/hep-th/9708005}{{\tt hep-th/9708005}}.

\bibitem{fgpwTwo}
D.~Z. Freedman, S.~S. Gubser, K.~Pilch, and N.~P. Warner, ``Continuous
  distributions of d3-branes and gauged supergravity,''
  \href{http://xxx.lanl.gov/abs/hep-th/9906194}{{\tt hep-th/9906194}}.

\bibitem{DuffLiu}
M.~J. Duff and J.~T. Liu, ``Anti-de Sitter black holes in gauged N = 8
  supergravity,'' \href{http://xxx.lanl.gov/abs/hep-th/9901149}{{\tt
  hep-th/9901149}}.

\bibitem{Myers}
A.~Chamblin, R.~Emparan, C.~V. Johnson, and R.~C. Myers, ``Charged AdS black
  holes and catastrophic holography,''
  \href{http://xxx.lanl.gov/abs/hep-th/9902170}{{\tt hep-th/9902170}}.

\bibitem{SethiSusskind}
S.~Sethi and L.~Susskind, ``Rotational invariance in the M(atrix) formulation
  of type IIB theory,'' {\em Phys. Lett.} {\bf B400} (1997) 265--268,
  \href{http://xxx.lanl.gov/abs/hep-th/9702101}{{\tt hep-th/9702101}}.

\bibitem{SeibergSixteen}
N.~Seiberg, ``Notes on theories with 16 supercharges,'' {\em Nucl. Phys. Proc.
  Suppl.} {\bf 67} (1998) 158,
  \href{http://xxx.lanl.gov/abs/hep-th/9705117}{{\tt hep-th/9705117}}.

\bibitem{russo}
J.~G. Russo, ``New compactifications of supergravities and large N QCD,''
  \href{http://xxx.lanl.gov/abs/hep-th/9808117}{{\tt hep-th/9808117}}.

\bibitem{witHolTwo}
E.~Witten, ``Anti-de Sitter space, thermal phase transition, and confinement in
  gauge theories,'' {\em Adv. Theor. Math. Phys.} {\bf 2} (1998) 505,
  \href{http://xxx.lanl.gov/abs/hep-th/9803131}{{\tt hep-th/9803131}}.

\bibitem{minConfine}
J.~A. Minahan, ``Asymptotic freedom and confinement from type 0 string
  theory,'' \href{http://xxx.lanl.gov/abs/hep-th/9902074}{{\tt
  hep-th/9902074}}.

\bibitem{gDil}
S.~S. Gubser, ``Dilaton driven confinement,''
  \href{http://xxx.lanl.gov/abs/hep-th/9902155}{{\tt hep-th/9902155}}.

\bibitem{KehagSfet}
A.~Kehagias and K.~Sfetsos, ``On running couplings in gauge theories from type
  IIB supergravity,'' \href{http://xxx.lanl.gov/abs/hep-th/9902125}{{\tt
  hep-th/9902125}}.

\bibitem{gog}
D.~J. Gross and H.~Ooguri, ``Aspects of large N gauge theory dynamics as seen
  by string theory,'' {\em Phys. Rev.} {\bf D58} (1998) 106002,
  \href{http://xxx.lanl.gov/abs/hep-th/9805129}{{\tt hep-th/9805129}}.

\bibitem{Uranga}
A.~M. Uranga, ``Brane configurations for branes at conifolds,''
  \href{http://xxx.lanl.gov/abs/hep-th/9811004}{{\tt hep-th/9811004}}.

\end{thebibliography}\endgroup
\bibliographystyle{ssg}

\end{document}